\documentclass[paper,11 pt]{article}

\usepackage{etoolbox}
\makeatletter

\apptocmd\@specialoutput{\global\holdinginserts\z@}

\makeatother

\usepackage{cite}
\usepackage{epsfig,amsfonts}
\usepackage{graphicx}
\usepackage{caption}
\usepackage{subcaption}

\usepackage{amsthm,amssymb}

\usepackage{hhline}
\usepackage{braket}
\usepackage{upgreek}
\usepackage[ruled,vlined]{algorithm2e}
\usepackage[linktocpage=true]{hyperref}
\usepackage[dvipsnames]{xcolor}
\usepackage{musicography}
\usepackage[normalem]{ulem}
\definecolor{purple_nice}{rgb}{0.4,0.2,0.7}
\definecolor{fuel_blue}{RGB}{42,162,185}
\definecolor{YInMn_blue}{RGB}{46, 80, 144}
\definecolor{ultramarine}{RGB}{63, 0, 255}
\definecolor{KLEIN_blue}{rgb}{0, 0.18, 0.65}
\hypersetup{
    colorlinks=true,
    linkcolor=ultramarine,
    citecolor=ultramarine,
    filecolor=fuel_blue,
    urlcolor=ultramarine,
}

\usepackage[T1]{fontenc}
\usepackage{amsmath}
\usepackage{mdframed}
\usepackage{graphics}
\usepackage{amsfonts}
\usepackage{amssymb}
\usepackage[multiple]{footmisc}
\usepackage{bm}
\usepackage{amsthm}
\usepackage{slashed}
\usepackage{enumerate,enumitem}
\usepackage{amsbsy}
\usepackage{bbm}
\usepackage{mathrsfs}
\usepackage{url}
\usepackage{bbm} 
\usepackage{color}
\usepackage{framed}
\usepackage{caption}
\usepackage{float}
\usepackage{cancel}
\usepackage{hyperref}
\usepackage{fnpct}
\usepackage{comment}
\newcommand{\aux}{\mathcal{H}}
\newcommand{\tr}{\mathrm{tr}}

\newcommand{\cP}{\mathcal{P}}

\newcommand{\cL}{\mathcal{L}}

\newcommand{\SM}{S_\mathrm{M}}
\newcommand{\SMt}{S_{\mathrm{M},\ta}}
\newcommand{\cR}{\mathcal{R}}

\newcommand{\Bx}{\mathbf{x}}

\newcommand{\Bg}{\mathbf{g}}
\newcommand{\Bh}{\mathbf{h}}
\newcommand{\BT}{\mathbf{T}}

\newcommand{\BR}{\bm{\mathcal{R}}}
\newcommand{\BJ}{\mathbf{J}}

\newcommand{\BD}{\mathbf{D}}
\newcommand{\BP}{\mathbf{P}}

\newcommand{\BF}{\mathbf{F}}

\newcommand{\dd}{\mathrm{d}}

\newcommand{\ta}{\tau}

\newcommand{\var}{\tau}
\newcommand{\vargrav}{\kappa}

\newcommand{\D}{d}

\newcommand{\TTb}{\mathrm{T}\overline{\mathrm{T}}}

\numberwithin{equation}{section}

\begin{document}
\begin{titlepage}

\title{{\huge \bf  Gravity and $\TTb$ flows in higher dimensions}}
\author{Tommaso Morone$^{1\musFlat}$, Stefano Negro$^{2,3 \musNatural}$ and Roberto Tateo$^{1\musSharp}$\\[0.3cm]}
\date{\small{$^1$ \textit{Dipartimento di Fisica, Università di Torino and \\ INFN, Sezione di Torino, Via P. Giuria 1, 10125, Torino, Italy} \\[0.2cm]$^2$ \textit{C.N. Yang Institute for Theoretical Physics, State University of New York,\\ Stony Brook, NY 11794-3840, U.S.A.} \\
[0.2cm]$^3$ \textit{Department of Mathematics, University of York,\\ Heslington, York YO10 5DD, United Kingdom}\\[0.3cm]
$^{\musFlat}$\texttt{\href{mailto:tommaso.morone@unito.it}{tommaso.morone@unito.it},} $^{\musNatural}$\texttt{\href{mailto:stefano.negro@york.ac.uk}{stefano.negro@york.ac.uk},}\\ $^{\musSharp}$\texttt{\href{mailto:roberto.tateo@unito.it}{roberto.tateo@unito.it}}\\}}
\maketitle
\thispagestyle{empty}

\begin{abstract}

\noindent We study systems in arbitrary space-time dimensions where matter, deformed by 
$\mathrm{T}\overline{\mathrm{T}}$-like irrelevant operators, is coupled to gravity in the Palatini formalism. The dynamically equivalent perspective is investigated, wherein the deformation transitions from the matter action to the gravitational one or vice versa. This alternative viewpoint leads to the emergence of Ricci-based gravity theories, thus providing a high-dimensional generalisation of the well-known equivalence between two-dimensional $\mathrm{T}\overline{\mathrm{T}}$ deformations and coupling to Jackiw-Teitelboim gravity. 
This dynamical equivalence is examined within the framework of the recently introduced Lagrangian flow equation, which notably led to the discovery of a direct link between Nambu-Goto theory and $\mathrm{T}\overline{\mathrm{T}}$ in $d=2$, as well as significant insights into nonlinear electrodynamics models in $d=4$.
The investigation involves explicit examples in $d=4$ dimensions; it builds upon earlier research concerning the metric interpretation of $\mathrm{T}\overline{\mathrm{T}}$-like perturbations, incorporates and extends recent findings in the cosmology-related literature associated to the concept of reframing. We focus on scenarios where the resulting modified gravity theories manifest as Born-Infeld and Starobinsky types.
\end{abstract}

\vspace{35mm}

\end{titlepage}
\newpage
\tableofcontents
\thispagestyle{empty}
\newpage
\section{Introduction}
\label{sec:introduction}
The $\TTb$ deformation of classical and quantum field theories in $d=2$ space-time dimensions \cite{Cavaglia:2016oda, Smirnov:2016lqw} has provided remarkable insights into the topology and geometry of the space of field theories, as well as allowing exact calculations of physical quantities related to the deformed models. In two dimensions, $\TTb$ flows are triggered by the operator
\begin{equation}\label{definition_intro}
     \TTb = \frac{1}{2}\left(\tr\left[\BT\right]^2-\,\tr\left[\BT^2\right]\right)= \det[\BT]\,,
\end{equation}
where $\BT$ denotes the two-dimensional stress-energy tensor of the theory. Although the $\TTb$ operator is irrelevant, it was shown that the local operator \eqref{definition_intro} is well
defined at a quantum level \cite{Zamolodchikov:2004ce}, and the flow it generates preserves many of the symmetries of the seed (i.e., undeformed) theory, including integrability. This last property is a feature of a much larger class of deformations, called double current deformations \cite{Dubovsky:2023lza}, and encompassing all the recent two-dimensional generalisations of the $\TTb$ deformation, such as the $\mathrm{J}\overline{\mathrm{T}}$ \cite{Guica:2017lia, Guica:2021pzy}, $\TTb_s$ \cite{Conti:2019dxg}, generalised $\TTb$ \cite{Hernandez-Chifflet:2019sua, Doyon:2021tzy, Doyon:2023bvo}, and CDD \cite{Camilo:2021gro,Cordova:2021fnr,He:2023obo} deformations. Moreover, many links have been observed with several topics in theoretical physics, such as string theory \cite{Baggio:2018gct, Dei:2018jyj, Chakraborty:2019mdf, Callebaut:2019omt}, holography \cite{McGough:2016lol, Giveon:2017nie, Gorbenko:2018oov, Kraus:2018xrn, Hartman:2018tkw, Guica:2019nzm, Jiang:2019tcq,Jafari:2019qns,Griguolo:2021wgy, Fichet:2023xbu}, random geometries \cite{Cardy:2018sdv}, out-of-equilibrium conformal field theory \cite{Medenjak:2020ppv, Medenjak:2020bpe}, the generalised hydrodynamics (GHD) approach \cite{Doyon:2021tzy, Cardy:2020olv, Doyon:2023bvo}, and quantum gravity \cite{Dubovsky:2017cnj, Dubovsky:2018bmo, Tolley:2019nmm, Iliesiu:2020zld, Okumura:2020dzb, Ebert:2022ehb, Bhattacharyya:2023gvg}. In particular, it was shown that any $\TTb$-deformed two-dimensional field theory is dynamically equivalent to its associated seed theory coupled to a topological theory of gravity \cite{Dubovsky_2017} which, on the plane, almost looks like Jackiw-Teitelboim gravity \cite{Teitelboim:1983ux,Jackiw:1984je}. In other words, denoting the seed theory by $S_{\mathrm{M}}$ and the corresponding $\TTb$ deformed theory by $S_{\mathrm{M},\ta}$, the following equivalence holds:
\begin{equation}\label{dubovsky}
    S_{\mathrm{M},\ta} \simeq  S_{\mathrm{M}} +    \int \mathrm{d}^2\Bx \sqrt{-g}\left(\varphi R - \Lambda_2\right),
\end{equation}
where vacuum energy $\Lambda_2$ is related to the $\TTb$ coupling parameter $\tau$ by $\ta \propto \Lambda_2^{-1}$. Notice that equation \eqref{dubovsky} provides a complete, quantum and non-perturbative definition of the $\TTb$ deformed theories along the whole flow.  
In addition, a main motivation for the current work stems from the observation that in $d=2$, a $\TTb$ deformation can be interpreted as a field-dependent local coordinate transformation that links the original model to its deformed version \cite{Dubovsky_2017, Conti:2018tca}. 

Generalisations of the $\TTb$ flow to higher dimensions have been introduced and studied in various works \cite{Taylor:2018xcy, Conti_2022, Bonelli:2018kik, Conti_Iannella, Cardy:2018sdv, Babaei-Aghbolagh:2020kjg, Hou:2022csf}, at least at the classical level. These investigations, alongside the introduction of the so-called  Modified Maxwell (ModMax) theory \cite{Bandos:2020jsw}, and the discovery that both Born-Infeld and ModMax arise from Maxwell theory through a Lagrangian flow involving $\TTb$-type composite fields \cite{Conti_Iannella, Babaei-Aghbolagh:2020kjg, Ferko:2022iru}, have sparked a revival of interest in nonlinear electrodynamics \cite{Bandos:2020jsw, Sorokin:2021tge, Lechner:2022qhb, Ferko:2019oyv}. Moreover, the fact that the corresponding deforming operators are constructed solely in terms of invariants built from the stress-energy tensor hints at the natural connection with General Relativity and modified gravity models, which will be discussed shortly.

Almost in parallel, the study of modified theories of gravity has gained substantial interest in cosmology. In particular, Born-Infeld-inspired minimal extensions of General Relativity allowed to reproduce  non-trivial gravitational dynamics while generating non-singular cosmologies \cite{Vollick:2003qp, Banados:2008rm, Banados:2008fj,Banados_2010, jimenez_2018, Olmo:2020fnk, Guerrero:2020azx, Nascimento:2019qor, Afonso:2019fzv, Pani:2012qb, Banerjee:2021auy, Olmo:2013gqa, Pereira:2023bxt}. Such theories, along with the more general Palatini-like theories of gravity, have been shown to admit dynamically equivalent Einstein-frame representations \cite{Magnano_1990, Afonso:2018hyj, Olmo_2022}, where the gravitational sector resembles the standard one from General Relativity and the ultraviolet corrections are transferred to the matter content of the theory.

This work investigates generalizations of \eqref{dubovsky} for $\TTb$-like flows in space-time dimensions $d>2$. To be more precise, denoting by $S_{\mathrm{G},\vargrav}$ a modified theory of gravity, depending on some mass scale $m_{\mathrm{G}}\propto \vargrav^{-\frac{1}{2}}$, and by $S_{\mathrm{M},\tau}$ a matter action depending on a $\TTb$-like flow parameter $\tau$, we will study explicit examples in $d=4$ within the following class of dynamical equivalences:
\begin{equation}
S_{\mathrm{G},\tau + \Delta\ta} + S_{\mathrm{M},\tau} \simeq S_{\mathrm{G},\tau} + S_{\mathrm{M},\tau+\Delta\ta}.
\label{eq:conj}
\end{equation}
In other words, we will provide non-trivial examples where modified gravity theories are coupled to $\TTb$-like deformed models, wherein the gravitational mass scale and the matter flow parameter are exchangeable on-shell, meaning once the equations of motion for the gravitational degrees of freedom have been enforced.

The remainder of this work is organised as follows. Section \ref{section2} reviews and extends recent findings regarding a class of $\TTb$-like deformations,  quadratic in the components of the stress-energy tensor. This class of Lagrangian perturbations will play a central role in our analysis. A significant mathematical simplification arises in $d=4$ and in examples that have historically garnered considerable attention. Section \ref{section2.1} illustrates how these models can be dressed or, in some cases, recovered through a metric transformation. The specific models that arise include self-dual theories of nonlinear electrodynamics, their Proca-like generalizations with generic interacting potentials, and models that coincide with or resemble those investigated in the context of inflationary cosmology. Section \ref{section3} introduces theories of gravity in the Palatini formalism, where the connection and the metric tensor are regarded as independent dynamical fields. Their dynamically equivalent representations are discussed, and metric deformations as well as modifications of the matter sectors naturally emerge within this framework. Section \ref{section4} provides a general scheme for linking Palatini theories of gravity with suitably crafted $\TTb$-like deformations in arbitrary dimensions. Explicit examples are worked out in four space-time dimensions, where algebraic simplifications allow for analytic solutions. We show the existence of dualities between Starobinsky gravity and trace-squared deformations, as well as between Eddington-inspired Born-Infeld gravity and $\TTb$-like deformations of Abelian gauge theories. Section \ref{section_final} examines stress-tensor flows triggered by arbitrary operators, and their relation to on-shell gravity flows. Appendix \ref{sec:dofs} provides some additional comments regarding the Jordan and Einstein frame representations of $f(\cR)$ gravity theories. Finally, Appendix \ref{unif_frame} analyses Modified Eddington-inspired Born-Infeld gravity, a family of Palatini theory that incorporates both Starobinsky and Eddington-inspired Born-Infeld gravity within a unified framework.

\section{$\TTb$-like deformations in arbitrary dimensions}\label{section2}

We denote as
\begin{equation}
    S_\mathrm{M} = \int \mathrm{d}^d\Bx \sqrt{-g} \,\cL_\mathrm{M}\,,\quad g:= \det\left[g_{\mu\nu}\right]
\end{equation}
a generic covariant matter action in $d$-dimensional space-time, where $\cL_\mathrm{M}$ is the associated Lagrangian density, which depends on the space-time coordinates $\Bx = \left(x^\mu\right)_{\mu \in \left\{0,\dots,d-1\right\}}$ through a generic collection of $N$ matter fields $\left\{\Phi_I\right\}_{I \in\{1, \ldots, N\}}$ and their higher-order derivatives. The field content of the theory is arbitrary unless otherwise stated. Indexes of tensors are lowered and raised using the metric $g_{\mu\nu}$ and its inverse $g^{\mu\nu}$, respectively, and repeated indexes are summed according to the Einstein notation. In this paper we focus on the family of deformations proposed in \cite{Conti_2022}, where the following flow equation for the matter sector has been considered:
\begin{equation}\label{deformedaction}
    \frac{\partial S_{\mathrm{M},\ta}}{\partial \ta} = \int \mathrm{d}^d\Bx \sqrt{-g}\,\mathcal{O}_{d,\ta}^{[a,b]}\,,\quad  S_{\mathrm{M},\ta_0}:= S_{\mathrm{M}}\,.
\end{equation}
The deforming operator in \eqref{deformedaction} is defined as
\begin{equation}\label{deformingoperator}
    \mathcal{O}_{d,\ta}^{[a,b]} := \frac{1}{d}\left(a\,\tr\left[\BT_\ta\right]^2-b\,\tr\left[\BT^2_\ta\right]\right),\quad a,\,b \in \mathbb{R}\,,\; d\geq 2\,,
\end{equation}
where $\ta$ is the flow parameter, and $\ta_0$ is a fixed point in correspondence to which the seed matter theory $S_\mathrm{M}$ lives. Here $\mathbf{T}_\tau=\left(g^{\mu\alpha}T_{\tau, \alpha\nu}\right)_{\mu, \nu \in\{0, \ldots, d-1\}}$ is a $d\times d$ dimensional matrix, where $T_{\ta,\mu\nu}$ are the components of the symmetric Hilbert stress-energy tensor
\begin{equation}
    T_{\ta,\mu\nu}=\frac{-2}{\sqrt{-g}} \frac{\delta S_{\mathrm{M},\ta}}{\delta g^{\mu \nu}}.
\end{equation}
In two dimensions, since
\begin{equation}\label{ttbar}
    \mathcal{O}_{2,\ta}^{[1,1]} = \frac{1}{2}\left(\tr\left[\BT_\ta\right]^2-\,\tr\left[\BT^2_\ta\right]\right)= \det[\BT_\ta]\,,
\end{equation}
we recover the usual definition of $\mathrm{T}\overline{\mathrm{T}}$ deformations when setting $a=1,\, b=1$. It is important to stress that this paper is about classical field theories and, apart from the special case \eqref{ttbar}, it is not known how to make the composite field \eqref{deformingoperator} well-defined at the quantum level. 

\subsection{$\TTb$-like dressing and the metric approach}
\label{sec:dressing}

In the context of two-dimensional $\mathrm{T}\overline{\mathrm{T}}$ deformations, it is known that the undeformed action $\SM$, with underlying metric tensor $g_{\mu\nu}$, is equivalent to the deformed one $\SMt$ over some new background metric $h_{\var,\mu\nu}$, up to a term proportional to the deforming operator evaluated in the seed theory \cite{Ceschin:2020jto, Coleman_2019}:
\begin{equation}\label{2d_dressing}
 \left.  \left\{ \SM[g_{\mu\nu},\Phi_I] -\left(\ta-\ta_0 \right)\int \mathrm{d}^2\Bx\sqrt{-g} \det\left[\BT_{\ta_0}\right]\right\}\right|_{g=g(h)} = \SMt\left[h_{\var,\mu\nu}, \Phi_I\right].
\end{equation}
Equation \eqref{2d_dressing} is colloquially known as the $\mathrm{T}\overline{\mathrm{T}}$ dressing mechanism: the flow in the space of field theories is balanced by an adiabatic flow of the background metric, which evolves through $\ta$. The associated deformed metric is  obtained as \cite{Conti:2018tca}
\begin{equation}
    h_{\var,\mu\nu} = g_{\mu\nu} -2(\tau-\ta_0) \epsilon_{\mu \alpha} \epsilon_{\nu\beta} T_{\ta}^{\alpha\beta}  -(\ta-\ta_0)^2 \epsilon_{\mu \alpha} \epsilon_{\nu\beta}T_{\ta,\rho}^\alpha T_{\ta}^{\rho\beta} 
  \,.
\end{equation}
The metric approach to two-dimensional deformations makes it possible to trade Lagrangian flows with perturbations of the underlying structure of space-time. Given our goal of exploring whether analogous geometric interpretations extend to higher-dimensional stress-tensor flows, the dressing of bare theories becomes an indispensable mathematical tool.

We should then look for families of stress-tensor deformations that naturally integrate within the framework of geometric flows, for which one must be able to find the deformed metric $h_{\var,\mu\nu}$ such that the generalized version of the $\TTb$ dressing holds:
\begin{equation}\label{bare_deformation}
    \left.  \left\{\SM[g_{\mu\nu},\Phi_I] -{\left(\ta-\ta_0 \right)} \int \mathrm{d}^d\Bx\sqrt{-g} \, \mathcal{O}_{d,\ta_0}^{[a,b]}\right\}\right|_{g=g(h)}  = \SMt\left[h_{\var,\mu\nu}, \Phi_I\right].
\end{equation}
Notice that, as in \eqref{2d_dressing}, the bare deforming operator appearing in \eqref{bare_deformation} is evaluated in the seed theory, defined at $\tau=\tau_0$. For the left-hand side of \eqref{bare_deformation} to match the right-hand side at $\tau=\tau_0$, one must have
\begin{equation}
    h_{\ta_0,\mu\nu} = g_{\mu\nu}\,.
\end{equation}
Moreover, expanding both sides of \eqref{bare_deformation} at first perturbative order in $\ta$, it is straightforward to show that the family of deformed metric $h_{\var,\mu\nu}$ must adhere to the constraint \cite{Conti_2022}
\begin{equation}\label{metric_flow_2}
  \frac{\mathrm{d} h_{\var,\mu\nu}}{\mathrm{d} \var} = -\dfrac{4}{d}\,\widehat{T}_{\var,\mu\nu}\,.
\end{equation}
Here we introduced the auxiliary tensor
\begin{equation}\label{widehatT}
\widehat{T}_{\var,\mu\nu} := a\, \tr_\ta\left[\BT_\var\right]g_{\var,\mu\nu} - b\, T_{\var,\mu\nu}\,,
\end{equation}
with the notation $\tr_\ta$ referring to traces being taken with respect to the $\ta$-dependent metric $h_{\ta,\mu\nu}$. Equation \eqref{metric_flow_2} defines an adiabatic flow along the trajectory of dynamically equivalent actions associated with the $\TTb$-type perturbation. Once on-shell over the dynamical degrees of freedom of the theory, each ``equilibrium'' configuration corresponds to a field-dependent modification of the background metric tensor according to \eqref{metric_flow_2}. Despite the somewhat restrictive choice of admissible operators made in \eqref{deformingoperator}, it is, in fact, possible to argue that the dressing mechanism, in the simple form described by \eqref{bare_deformation}, can be realised if and only if the deformation is driven by operators that are quadratic functionals of the stress-energy tensor\footnote{This can be also understood perturbatively from the geometric perspective introduced by J. Cardy in \cite{Cardy:2018sdv}, which relies on the Hubbard-Stratonovich transformation to account for the metric deformation in the $\TTb$-deformed partition function. If the deforming operators were not quadratic functionals of the stress-energy tensor, the fluctuations about the saddle point would not be independent of the stress-energy tensor.}. The following section aims to examine whether it is possible -- at least in a few special cases -- to obtain the exact analytic expression for the metric  $h_{\ta,\mu\nu}$  associated with the $\TTb$-like dressing mechanism \eqref{bare_deformation}.
\subsection{Exact solutions to the metric flow}
Exact solutions for the metric flow \eqref{metric_flow_2} can be derived algorithmically, as detailed in \cite{Conti_2022}. The idea is to Taylor expand the deformed metric $h_{\var,\mu\nu}$ around $\var = \var_0$ as
\begin{equation}\label{taylor}
    h_{\var,\mu\nu} = \sum_{n=0}^\infty \frac{h^{(n)}_{\var_0,\mu\nu} }{n!}\left(\var-\var_0\right)^n,\qquad h^{(0)}_{\var_0,\mu\nu} = g_{\mu\nu}.
\end{equation}
The first coefficient $h^{(1)}_{\var_0,\mu\nu}$ descends trivially from \eqref{metric_flow_2}, yielding
\begin{equation}
     h^{(1)}_{\var_0,\mu\nu} = -\frac{4}{d}\,\widehat{T}_{\var_0,\mu\nu}\,.
\end{equation}
On the other hand, relying on the definition of $T_{\ta,\mu\nu}$, the partial derivative of the stress-energy tensor with respect to the flow parameter can be computed as
\begin{equation}\label{tvar}
    \frac{\partial T_{\ta,\mu\nu}}{\partial \ta} =-\frac{2}{\sqrt{-h}} \frac{\partial}{\partial\ta} \left(\frac{\delta \SMt}{\delta h_\ta^{\mu\nu}}\right) = -\frac{2}{d\sqrt{-h}}\frac{\delta}{\delta (h_\ta^{-1})^{\mu\nu}}\left(\sqrt{-h}\,T_{\ta,\alpha\beta}\widehat{T}_{\ta}^{\alpha\beta}\right)\,,
\end{equation}
where $h := \det[h_{\mu\nu}]$. Equation \eqref{tvar}, together with \eqref{widehatT}, allows us to compute the full variation
\begin{equation}\label{full}
    \frac{\mathrm{d} \widehat{T}_{\var,\mu\nu}}{\mathrm{d} \var} = -\dfrac{4}{d}\,\widehat{T}_{\var,\mu}^\alpha \widehat{T}_{\var,\alpha\nu} - \xi_\var \widehat{T}_{\var,\mu\nu} - \chi_\var h_{\var,\mu\nu}\,,
\end{equation}
where $\xi_\var$ and $\chi_\var$ are scalar functionals of the stress-energy tensor, respectively defined as
\begin{equation}
      \xi_\var = \frac{2}{d}\left(b-da\right)\tr_\ta \left[\BT_\var\right],\qquad \chi_\var = \frac{da-b}{d}\left(a\,\tr_\ta\left[\BT_\var\right]^2-b\,\tr_\ta\left[\BT^2_\var\right]\right).
\end{equation}
From equation \eqref{full}, the second coefficient of the Taylor expansion \eqref{taylor} is obtained as
\begin{equation}
    h^{(2)}_{\var_0,\mu\nu} = \left(\frac{4}{d}\right)^{\!2}\widehat{T}_{\var_0,\mu}^\alpha \widehat{T}_{\var_0,\alpha\nu} + \frac{4}{d}\,\xi_{\var_0} \widehat{T}_{\var_0,\mu\nu} + \frac{4}{d}\,\chi_{\var_0} g_{\mu\nu} \,.
\end{equation}
The higher order coefficients have been computed recursively in \cite{Conti_2022}, by considering the general expression 
\begin{equation}
    h_{\ta,\mu \nu}^{(n)}=c_0^{(n)} h_{\ta,\mu \nu}+\sum_{k=1}^n c_k^{(n)} \widehat{T}_{\ta, \mu \nu}^k, \quad \forall n \geq 1\,.
\end{equation}
Here $\{c_k^{(n)}\}_{k \in\{1, \ldots, n\}}$ are polynomials in the variables $\xi_\ta$ and $\chi_\ta$ with real coefficients. In this fashion, the computation of each $h_{\ta,\mu \nu}^{(n)}$ can be reduced to the computation of the coefficients $\{c_k^{(n)}\}_{k \in\{1, \ldots, n\}}$, bound by the recurrence relations
\begin{equation}
\label{eq:coeffrecursive}
    \begin{cases}
    \displaystyle{c_0^{(n+1)}=-\frac{\dd c_0^{(n)}}{\dd\var}-\chi_\var c_1^{(n)}}\vspace{1mm}\\
    \displaystyle{c_{k}^{(n+1)}=-\frac{4}{\D}c_{k-1}^{(n)}-\frac{\dd c_{k}^{(n)}}{\dd\var}-k\xi_\var c_{k}^{(n)}-\left(k+1\right)\chi_\var c_{k+1}^{(n)} \;,\quad 1\leq k\leq n-1}\vspace{1mm}\\
    \displaystyle{c_n^{(n+1)}=-\frac{4}{\D}c_{n-1}^{(n)}-\frac{\dd c_{n}^{(n)}}{\dd\var}-n\xi_\var c_{n}^{(n)}}\vspace{1mm}\\
    \displaystyle{c_{n+1}^{(n+1)}=-\frac{4}{\D}c_{n}^{(n)}}
    \end{cases}\;
\end{equation}
The recurrence relations \eqref{eq:coeffrecursive} can be implemented in a Mathematica notebook, with the first $n=100$ coefficients of the series $\{c_k^{(n)}\}_{k \in\{1, \ldots, n\}}$ obtained in less than a minute on a standard machine.
%
\section{Deformed actions from the change of metric}\label{section2.1}
In this section, we will focus our attention on a few notable examples where the metric series expansion exhibits low-order truncations.
\subsection{Trace-squared deformations}
Setting $b=0$ in \eqref{deformingoperator} we obtain trace-squared deformations in arbitrary dimensions, with characteristic flow equation
\begin{equation}\label{flow_tr_sq_d}
   \frac{\partial S_{\mathrm{M},\var}}{\partial \var} = \frac{a}{d}\int \mathrm{d}^4\Bx \sqrt{-g}\,\tr\left[\BT_\var\right]^2\,.  
\end{equation}
It is important to note that $a$ is not an independent parameter, as it can be set to 1 by rescaling the flow parameter as $\tau \rightarrow \tau/a$. From \eqref{metric_flow_2}, the associated flow in the space of metrics is given by
\begin{equation}\label{trace_squared_metric_flow}
   \frac{\mathrm{d}h_{\ta,\mu\nu}}{\mathrm{d}\ta} = -\frac{4a}{d}\tr_\ta[\BT_\ta] h_{\ta,\mu\nu}\,.  
\end{equation}
It is possible to integrate equation \eqref{trace_squared_metric_flow} via the recurrence relations \eqref{eq:coeffrecursive}, which yield\footnote{It must be noted that equation \eqref{recurrence_trace_squared} does not stem from analytic computations; nevertheless, its validity has been substantiated through explicit verification across arbitrary space-time dimensions, extending up to the 100th order in the Taylor expansion.}
\begin{equation}\label{recurrence_trace_squared}
    h_{\ta,\mu\nu} = \left(1-a(\ta-\ta_0)\tr[\BT_{\ta_0}]\right)^{\frac{4}{d}}g_{\mu\nu}\,.
\end{equation}
In other words, trace-squared deformations amount to a field-dependent Weyl rescaling of the background metric. Note that the ${4}/{d}$ exponent appearing in \eqref{recurrence_trace_squared} does not stand as an exclusive hallmark of trace-squared deformations, as we will show in the following section. Ultimately, its presence is due to the ${4}/{d}$ factor in \eqref{metric_flow_2}, and it anticipates the special role played by $\TTb$-like deformations in four space-time dimensions. Once the full form of the deformed metric is known, the $\TTb$-like dressing machinery can be put to work over physically relevant theories: specifically, for a given matter seed theory $S_\mathrm{M}$, its associated deformed theory $\SMt$ can be computed as
\begin{equation}\label{trsqdressing}
  \left.  \left\{\SM[g_{\mu\nu},\Phi_I] -{\frac{a\left(\ta-\ta_0 \right)}{d}} \int \mathrm{d}^d\Bx\sqrt{-g} \,\tr[\BT_{\ta_0}]^2\right\}\right|_{g=g(h)}  = \SMt\left[h_{\var,\mu\nu}, \Phi_I\right],
\end{equation}
with $h_{\ta,\mu\nu}$ given by \eqref{recurrence_trace_squared}. There are two interesting scenarios\footnote{There is, actually, a third possibility: a trace-squared deformed one-dimensional field theory (i.e., a mechanical system evolving in time) is equivalent to a metric deformation (i.e., a change in the clock of the system) through a Weyl rescaling characterized by a quartic dependency on the energy of the underlying seed theory.} in which the expression \eqref{recurrence_trace_squared} assumes a particularly simple structure: namely, $d=4$, where the deformation of the metric is linear in $\ta$, and $d=2$, where the deformation is quadratic in the flow parameter.
\subsubsection*{Trace-squared deformations in $d=4$ space-time} 
As an instructive example, we start by examining the effects of trace-squared deformations on the action that describes a single interacting boson in $d=4$. To this end, we introduce the following action at $\ta_0=0$:
\begin{equation}\label{boson}
    S_{V}\left[g_{\mu\nu},\phi\right] = \int \dd^4 \Bx \sqrt{-g}\left[\frac{1}{2}\partial_{\mu}\phi \partial^\mu\phi - V(\phi)\right].
\end{equation}
Here $V$ denotes an arbitrary Lorentz-invariant potential, and for notational convenience, we introduce the symmetric, metric-independent quantity
\begin{equation}
    X_{\mu\nu} = \partial_\mu\phi\partial_\nu\phi\,,
\end{equation}
whose trace we denote by $X = g^{\mu\nu}X_{\mu\nu}$. The trace of the stress-energy tensor associated with the seed theory is given by
\begin{equation}
    \tr[\BT_0] = X-4V\,,
\end{equation}
so that, by equation \eqref{recurrence_trace_squared}, the deformed metric is
\begin{equation}\label{h-g-4dts}
    h_{\ta,\mu\nu} = \left(1-a\ta\tr[\BT_{0}]\right)g_{\mu\nu} = \left(1-a\ta\left(X-4V\right)\right)g_{\mu\nu}\,.
\end{equation}
The next step of the $\TTb$-like dressing requires inverting equation \eqref{h-g-4dts}, expressing the metric $g_{\mu\nu}$ as a function of $h_{\ta,\mu\nu}$. Before we do that, notice that the kinetic term $X$ appearing in \eqref{h-g-4dts} implicitly contains the background metric $g_{\mu\nu}$, hidden inside the trace over space-time derivatives. For this reason, we introduce the term
\begin{equation}
     X_\ta := (h_\ta^{-1})^{\mu\nu}X_{\mu\nu} = (h_\ta^{-1})^{\mu\nu}\partial_\mu\phi\partial_\nu\phi\,,
\end{equation}
which is related to the original kinetic term of the undeformed theory via
\begin{equation}\label{xta-x}
    X_\ta = \frac{X}{1-a\ta\left(X-4V\right)}\,,
\end{equation}
or, explicitly solving \eqref{xta-x} for $X$,
\begin{equation}\label{newx}
    X = \frac{X_\ta\left(1+4a\ta V\right)}{1+a\ta X_\ta}\,.
\end{equation}
At this point, we have all the necessary ingredients to dress up the action \eqref{boson} with the trace-squared deformation. First, we subtract the deforming operator computed at $\ta=\ta_0$ (in this case, we have fixed $\ta_0=0$):
\begin{equation}\label{subinto}
    S_V\left[g_{\mu\nu},\phi\right] - \frac{a\ta}{4}\int \mathrm{d}^4\Bx \sqrt{-g}\,\tr\left[\BT_0\right]^2 = S_V\left[g_{\mu\nu},\phi\right] - \frac{a\ta}{4}\int \mathrm{d}^4\Bx \sqrt{-g}\left(X-4V\right)^2\,.
\end{equation}
Next, we substitute $g=g(h_\ta)$ in \eqref{subinto} as determined by \eqref{h-g-4dts}. Taking into account the transformation rule \eqref{newx}, and keeping in mind that in four space-time dimensions the metric determinant in \eqref{boson} transforms as
\begin{equation}
    \sqrt{-g} = \frac{\sqrt{-h}}{(1-a\ta\tr[\BT_0])^2} = \frac{\sqrt{-h}}{\left(1-a\ta\left(X-4V\right)\right)^2}\,,
\end{equation}
we obtain the deformed action
\begin{equation}\label{deformed_free_boson_new}
\begin{split}
  S_{V,\ta}[h_{\ta,\mu\nu},\phi] &= \left.\left\{ S_V\left[g_{\mu\nu},\phi\right] - \frac{a\ta}{4}\int \mathrm{d}^4\Bx \sqrt{-g}\,\tr\left[\BT_0\right]^2  \right\}\right|_{g=g(h)} \\&= \int \mathrm{d}^4\Bx\sqrt{-h_\ta}\left(\frac{2X_\ta+a\ta X_\ta^2}{4\left(1+4 a \ta V\right)}- \frac{V}{1+4a\ta V}\right)\,.
   \end{split}
\end{equation}
As a final check, notice that the trace of the deformed stress-energy tensor -- here computed over the metric structure induced by $h_{\ta,\mu\nu}$ -- is given by
\begin{equation}
 \tr_\ta\left[\BT_\ta\right] = \frac{X_\ta-4 V}{1+4 a \tau  V}\,.
\end{equation}
If we now regard the metric $h_{\ta,\mu\nu}$ as fixed, and compute the derivative of \eqref{deformed_free_boson_new} with respect to $\ta$, we obtain
\begin{equation}
    \frac{\partial}{\partial\ta}S_{V,\ta}[h_{\ta,\mu\nu},\phi] = \frac{a}{4}\int \dd^4 \Bx\sqrt{-h_\ta}\left(\frac{X_\ta-4V}{1+4 a \tau  V}\right)^2 = \frac{a}{4}\int \dd^4 \Bx\sqrt{-h_\ta}\,\tr_\ta\left[\BT_\ta\right]^2\,.
\end{equation}
Since there is no harm in simply relabelling $h_{\ta,\mu\nu}$ as $g_{\mu\nu}$, we have shown that the action 
\begin{equation}\label{def_sc_1}
     S_{V,\ta}[g_{\mu\nu},\phi] = \int \mathrm{d}^4\Bx\sqrt{-g}\left(\frac{2X+a\ta X^2}{4\left(1+4 a \ta V\right)}- \frac{V}{1+4a\ta V}\right)
\end{equation}
satisfies the flow \eqref{flow_tr_sq_d} in $d=4$, with \eqref{boson} as the associated seed theory in $\ta=0$. Further generalisations in four dimensions can account for the introduction of multiple scalar fields $\left\{\phi_I\right\}_{I \in \left\{1,\dots,N\right\}}$. In this case, the seed theory
\begin{equation}
   S_{V}^{(N)} \left[g_{\mu\nu},\phi_I\right] =\int \dd^4 \Bx \sqrt{-g}\left(\frac{1}{2}\partial_\mu\phi_I\partial^\mu\phi^I - V\right) \,,
\end{equation}
where implicit summation over the field index $I=1,\dots,N$ is understood, is deformed by the trace-squared deformation into
\begin{equation}
     S_{V,\ta} ^{(N)}\left[g_{\mu\nu},\phi_I\right]=\int \dd^4 \Bx \sqrt{-g}\left(\frac{2\partial_\mu\phi_I\partial^\mu\phi^I + a \ta \left(\partial_\mu\phi_I\partial^\mu\phi^I\right)^2}{4(1+4a\ta V)} - \frac{V}{1+4a\ta V}\right)\,.
\end{equation}
Analogue procedures can be implemented when the matter Lagrangian exhibits even more intricate structures. For example, consider the action describing self-interacting scalar fields minimally coupled to Maxwell's electrodynamics:
\begin{equation}\label{sqed}
    S[g_{\mu\nu},\Phi,A_\mu] = \int \dd^4 \Bx \sqrt{-g}\left[\mathcal{D}_\mu \Phi \left(\mathcal{D}^\mu\Phi\right)^\star -V\left(\Phi\Phi^\star\right) -\frac{1}{4}F_{\mu\nu}F^{\mu\nu}\right].
\end{equation}
Here, $F_{\mu\nu} = \partial_\mu A_\nu - \partial_\nu A_\mu$ is the field strength for the Abelian gauge field $A_{\mu}$, $\Phi$ is a complex scalar field, $V$ is an arbitrary potential, and the matter-radiation coupling is realised through the introduction of the covariant derivative $\mathcal{D}_\mu := \partial_\mu - i A_\mu$. One can explicitly check that \eqref{sqed} is deformed by the trace-squared perturbation into
\begin{equation}
    S_\ta[g_{\mu\nu},\Phi,A_\mu] = \int \dd^4 \Bx \sqrt{-g}\left(\frac{\mathcal{D}_\mu \Phi \left(\mathcal{D}^\mu\Phi\right)^\star + a\ta \left[\mathcal{D}_\mu \Phi \left(\mathcal{D}^\mu\Phi\right)^\star\right]^2-V}{1+4a\ta  V}-\frac{1}{4}F_{\mu\nu}F^{\mu\nu}\right).
\end{equation}
There is an intuitive reason for this. First, we notice that the stress tensor associated with the kinetic sector of the electromagnetic theory in \eqref{sqed} is traceless in $d=4$. Since the deformed theory is constructed from the seed action by recursively incorporating functionals of the trace, the $F_{\mu\nu}F^{\mu\nu}$ term remains untouched, as it receives no contributions. On the other hand, the kinetic sector of the scalar field theory matches the one obtained in \eqref{def_sc_1}, providing that we appropriately rescale $\Phi \to \Phi/\sqrt{2}$ to replicate the initial normalisation, and substitute $\partial_\mu \to \mathcal{D}_{\mu}$ as required by the minimal coupling prescription.

\subsubsection*{Trace-squared deformations in $d=2$ space-time}
In two space-time dimensions, the metric deformation associated with trace-square deformations is quadratic in $\ta$. Again, we start by considering a single interacting scalar field as our seed theory at $\ta_0=0$, whose action is given by:
\begin{equation}\label{boson2d}
    S_{V}\left[g_{\mu\nu},\phi\right] = \int \dd^2 \Bx \sqrt{-g}\left[\frac{1}{2}\partial_{\mu}\phi \partial^\mu\phi - V(\phi)\right].
\end{equation}
The trace of the stress-energy tensor associated with \eqref{boson2d} is easily computed as 
\begin{equation}
    \tr[\BT_0] = -2V\,.
\end{equation}
The deformed background metric is then related to the original one by
\begin{equation}\label{h-g-2dts}
    h_{\ta,\mu\nu} = \left(1-a\ta\tr[\BT_{0}]\right)^2 g_{\mu\nu} = \left(1+2a\ta V\right)^2 g_{\mu\nu}\,.
\end{equation}
Recalling that in $d=2$ the metric determinants transform under \eqref{h-g-2dts} as
\begin{equation}
    \sqrt{-g} = \frac{\sqrt{-h}}{1-a\ta\tr[\BT_0]} = \frac{\sqrt{-h}}{1+2a\ta V}\,,
\end{equation}
the action \eqref{boson2d} is dressed by the trace-squared deformation into
\begin{equation}\label{deformed_free_boson2_new}
\begin{split}
  S_{V,\ta}[h_{\ta,\mu\nu},\phi] &= \left.\left\{ S_V\left[g_{\mu\nu},\phi\right] - \frac{a\ta}{2}\int \mathrm{d}^2\Bx \sqrt{-g}\,\tr\left[\BT_0\right]^2  \right\}\right|_{g=g(h)} \\&= \int \mathrm{d}^2\Bx\sqrt{-h_\ta}\left(\frac{1}{2}X_\ta- \frac{V}{1+2a\ta V}\right)\,,
   \end{split}
\end{equation}
where again $X_\ta =h_\ta^{\mu\nu}\partial_\mu\phi\partial_\nu\phi$. Relabelling $h_{\ta,\mu\nu}$ as $g_{\mu\nu}$ in \eqref{deformed_free_boson2_new}, and regarding $g_{\mu\nu}$ as fixed, one can verify that the trace-squared flow \eqref{flow_tr_sq_d} is satisfied. Note that the kinetic term is unaffected by the deformation: this is analogous to the behaviour of pure Maxwell theory in $d=4$ space-time. In $d=2$, this phenomenon finds its ultimate origin in the intrinsic property of conformal field theories, wherein stress-energy tensors are inherently traceless, without the necessity for improvement procedures. It is also interesting to observe that, as $\ta\to\infty$, the deformed theory \eqref{deformed_free_boson2_new} becomes free.

\subsection{Deformations of Abelian gauge theories}
Assume that the matrix $\BT_{\var_0}$ is diagonalisable, i.e., there exists an invertible matrix $\BP$ and a diagonal matrix $\BD$ such that $\BT_{\var_0}=\BP\BD\BP^{-1}$. Moreover, assume that $\BT_{\var_0}$ has 2 (respectively, 1) independent eigenvalues of multiplicity $d/2$ (respectively, $\D$) if $\D$ is even (respectively, odd), namely
\begin{equation}
\label{eq:Ddiag}
\BD=
\begin{cases}
\text{diag}\,\bigl(\underbrace{\lambda_1,\dots,\lambda_1}_\text{$\frac{\D}{2}$-times},\underbrace{\lambda_2,\dots,\lambda_2}_\text{$\frac{\D}{2}$-times}\bigr) \;,\quad \D\in2\mathbb{N}+2 \\
\text{diag}\,\bigl(\underbrace{\lambda,\dots,\lambda}_\text{$\D$-times}\bigr) \;,\quad \D\in2\mathbb{N}+3\,.
\end{cases}\;
\end{equation}
Under the assumption \eqref{eq:Ddiag}, fixing $a=2/d$ and $b=1$ in \eqref{deformingoperator}, the coefficients of the Taylor expansion \eqref{taylor} can be computed as
\begin{equation}
    h_{\ta_0,\mu\nu}^{(n)}=(-1)^n\left(-\frac{4}{d}\right)_n \widehat{T}_{\ta_0, \mu \nu}^n\,,
\end{equation}
where we introduced the Pochhammer symbol defined as
\begin{equation}
    \left(x\right)_n = \frac{\Gamma(x+n)}{\Gamma(x)}\,.
\end{equation}
At least formally, the deformed metric can be then written as
\begin{equation}\label{def_gauge_metric}
    h_{\ta,\mu\nu} =\left[\left(g-\left(\ta-\ta_0\right) \widehat{T}_{\ta_0}\right)^{\frac{4}{d}}\right]_{\mu\nu}\,.
\end{equation}
As anticipated when discussing trace-squared deformations, the four-dimensional scenario assumes yet again a distinctive significance, as it makes \eqref{def_gauge_metric} linear in $\ta$. In arbitrary space-time dimensions, provided that the theory satisfies \eqref{eq:Ddiag}, the dressing equation \eqref{bare_deformation} takes the explicit form
\begin{equation}\label{gauge_dressing}
    \left.  \left\{\SM[g_{\mu\nu},\Phi_I] -{\frac{\left(\ta-\ta_0 \right)}{d}} \int \mathrm{d}^d\Bx\sqrt{-g} \left(\frac{2}{d}\tr\left[\BT_{\ta_0}\right]^2-\,\tr\left[\BT^2_{\ta_0}\right]\right)\right\}\right|_{g=g(h)}  = \SMt\left[h_{\var,\mu\nu}, \Phi_I\right]\,,
\end{equation}
with $h_{\ta,\mu\nu}$ provided by \eqref{def_gauge_metric}. It is interesting to notice that, due to the degeneracy properties of the stress-energy tensor \eqref{eq:Ddiag}, in even space-time dimensions one has
\begin{equation}\label{determinant_form_u1}
    \frac{1}{d}\left(\frac{2}{d}\tr\left[\BT_{\ta_0}\right]^2-\,\tr\left[\BT^2_{\ta_0}\right]\right) = \lambda_1\lambda_2 = \left(\det[\BT_{\ta_0}]\right)^{\frac{2}{d}}\,.
\end{equation}
Of course, the same statement holds in odd dimensions, under the identification $\lambda_1=\lambda_2=\lambda$. One naturally ponders which physically relevant theories adhere to the constraints delineated by \eqref{eq:Ddiag}. As it turns out, such theories are more common than one may expect:
\begin{itemize}
    \item in $d=2$, the condition \eqref{eq:Ddiag} does not constrain the stress-energy tensor which has, in general, 2 distinct eigenvalues. The family of stress-tensor flows reduces to
    \begin{equation}
       \frac{\partial S_{\mathrm{M},\ta}}{\partial \ta} = \frac{1}{2}\int \mathrm{d}^2\Bx \sqrt{-g}  \left(\tr\left[\BT_{\ta}\right]^2-\,\tr\left[\BT^2_{\ta}\right]\right) = \int \mathrm{d}^d\Bx \sqrt{-g} \,\det[\BT_\ta]\,,
    \end{equation}
    which reproduce the usual $\TTb$ deformations in two-dimensional spacetime;
    \item in $d=4$, as we shall soon discuss, Abelian gauge theories are characterised by the stress tensor degeneracy required by \eqref{eq:Ddiag} \cite{Ferko_2022}, and the $\TTb$-like flow reads
    \begin{equation}\label{4dflow_gauge}
       \frac{\partial S_{\mathrm{M},\ta}}{\partial \ta} = \frac{1}{4}\int \mathrm{d}^4\Bx \sqrt{-g}  \left(\frac{1}{2}\tr\left[\BT_{\ta}\right]^2-\,\tr\left[\BT^2_{\ta}\right]\right);
    \end{equation}
\end{itemize}
In addition, it was recently discovered that in the chiral two-form theories in $d=6$, the stress-energy tensor also displays eigenvalue degeneracy (see comments at the end of section 7 of \cite{Ferko:2024zth}).

\subsubsection*{A warm-up: from Maxwell to Maxwell-Born-Infeld}
Before we dive into the more general framework of $d=4$ Abelian gauge theories, it is instructive to analyse what happens when we take into account the simplest $U(1)$ gauge theory in four space-time dimensions. Namely, we focus on Maxwell's theory, which we take as our seed theory in $\ta_0=0$, and whose action is given by
\begin{equation}\label{max_action}
    S_{\mathrm{Max}}[g_{\mu\nu},A_\mu] = -\frac{1}{4}\int\dd^4 \Bx \sqrt{-g}F_{\mu\nu}F^{\mu\nu}\,.
\end{equation}
We compute the stress-energy tensor associated with \eqref{max_action} in its matrix form, which reads 
\begin{equation}\label{tmax_matrix}
    {\BT_{0} = \BF^2-\frac{1}{4} \tr\left[\BF^2\right]\mathbf{1}\,,}
\end{equation}
where $\mathbf{1}$ denotes the $4\times 4$ identity matrix, and we introduced $\BF=\left(g^{\mu\alpha}F_{\alpha\nu}\right)_{\mu, \nu \in\{0, \ldots, 3\}}$. Notice that $\BF$ is, by construction, antisymmetric. If we label as $\{\ell_n\}_{n\in\{0,\dots,3\}}$ the eigenvalues of $\BF$, since the matrix \eqref{tmax_matrix} is symmetric, it is always possible to find a matrix $\mathbf{P}$ such that $\BT_{0}=\BP\BD\BP^{-1}$, with $\BD$ given by
\begin{equation}
    \BD = \text{diag}\left(\frac{1}{4} \tr\left[\BF^2\right] - \ell_0^2, \dots, \frac{1}{4} \tr\left[\BF^2\right] - \ell_3^2\right):=\text{diag}\left(\lambda_0, \dots, \lambda_3\right) \,.
\end{equation}
Moreover, since the eigenvalues of any antisymmetric matrix are purely imaginary and come in complex conjugate pairs, we can take
\begin{equation}
\left\{\ell_0,\ell_1,\ell_2,\ell_3\right\} = \left\{-\ell_1,\ell_1,\ell_2,-\ell_2\right\}.
\end{equation}
Given that only the square of each $\ell_n$ contributes to the matrix $\BD$, we obtain
\begin{equation}\label{degeneracy_max}
  \BD = \text{diag}\left(\lambda_1,\lambda_1,\lambda_2,\lambda_2\right)\,.   
\end{equation}
This is enough to ensure that, under the $\TTb$-like flow \eqref{4dflow_gauge}, Maxwell's action \eqref{max_action} is dressed as 
\begin{equation}\label{max_dressing}
    \left.  \left\{S_{\mathrm{Max}}[g_{\mu\nu},A_\mu] +{\frac{\ta}{4}} \int \mathrm{d}^4\Bx\sqrt{-g}\,\tr\left[\BT^2_{0}\right]\right\}\right|_{g=g(h)}  = S_{\mathrm{Max},\ta}\left[h_{\var,\mu\nu}, \Phi_I\right]\,,
\end{equation}
where, according to \eqref{def_gauge_metric}, we take the deformed metric $h_{\ta,\mu\nu}$ to be
\begin{equation}\label{max_metric_def}
    h_{\ta,\mu\nu} = g_{\mu\nu} +\ta T_{0,\mu\nu}\,.
\end{equation}
Notice that the $\tr\left[\BT_{0}\right]^2$ term in \eqref{gauge_dressing} has disappeared from \eqref{max_dressing}, since the stress-energy tensor of the seed theory is traceless in $d=4$. In addition, it is worth pointing out that we can express equation \eqref{max_metric_def} in a more explicit way as follows:
\begin{equation}\label{explicit_max_mdef}
 {\Bh_\ta = \BJ^\dagger\Bg\, \BJ, \qquad   \BJ :=\sqrt{1-\frac{\tau}{4} \tr\left[\BF^2\right]}\,\mathbf{1} -i\sqrt{\ta}\, \BF \;,}
\end{equation}
where $\BJ$ is a $4\times 4$ square matrix with complex entries. The transformation of the metric as written in \eqref{explicit_max_mdef} can be easily understood as a field-dependent local change of coordinates, characterised by the Jacobian matrix $\BJ$. Consistently with \eqref{max_metric_def}, we also have,
\begin{equation}
{\BJ\BJ^\dagger =\BJ^\dagger \BJ = \mathbf{1}+\ta \BF^2-\frac{1}{4}\ta \tr\left[\BF^2\right]\mathbf{1}  =  \mathbf{1}+\ta \BT_{0}}\;.
\end{equation}
In line with our previous approach to scalar fields in the context of trace-squared deformations, we need to define new quantities in the $h_{\ta,\mu\nu}$ frame. In this case, it suffices to introduce the matrix $\mathbf{F}_\ta=\left(h^{\mu\alpha}F_{\alpha\nu}\right)_{\mu, \nu \in\{0, \ldots, 3\}}$, related to the original $\BF$ via
\begin{equation}
    \BF_\ta = \left(\BJ^{-1}\right)^\dagger\BF\left(\BJ^{-1}\right)\,.
\end{equation}
Finally, given that
\begin{equation}
    \sqrt{-h_\ta} = \det\left[\BJ\right] \sqrt{-g}\;,
\end{equation}
one can algebraically manipulate equation \eqref{max_dressing} to obtain an explicit expression for the deformed action:
\begin{equation}\label{mbi_action_h}
S_{\mathrm{Max},\ta}\left[h_{\var,\mu\nu}, A_\mu\right] =  \frac{1}{2\ta}\int\dd^4\Bx\sqrt{-h_\ta}\left(\sqrt{\det\left[\mathbf{1}+\sqrt{2 \tau}\BF_\ta\right]}-1\right)\,.    
\end{equation}
As is customary, one may relabel redundant quantities to yield a more succinct form for the action \eqref{mbi_action_h}: we restore the notation for a generic background metric $g_{\mu\nu}$, and set $2\ta = 1/\beta^2$, yielding 
\begin{equation}\label{mbi_action_standard}
S_{\mathrm{Max},\ta}\left[g_{\mu\nu}, A_\mu\right] =  \beta^2\int\dd^4\Bx\sqrt{-g}\left(\sqrt{\det\left[\mathbf{1}+\frac{1}{\beta}\BF\right]}-1\right)\,.   
\end{equation}
Here $\beta$ is simply regarded as a scale parameter. The action \eqref{mbi_action_standard} is the renowned Maxwell-Born-Infeld action, historically introduced in the 1930s to remove the divergences emerging due to the electron's self-energy in classical electrodynamics. In this theory, the maximum attainable value of the electric field is $\beta$, ensuring finite self-energy for point charges. Initially shadowed by the advent of renormalisation, the Born-Infeld theory regained popularity in the mid-80s as the leading term in the low-energy effective action of the open string theory expanded in powers of derivatives of gauge field strength \cite{Fradkin:1985qd}. It comes without surprise that the dressing mechanism \eqref{max_dressing} applied to Maxwell's theory reproduces the Maxwell-Born-Infeld theory, as it was shown in \cite{Conti_Iannella} that \eqref{mbi_action_standard} is indeed a solution to the four-dimensional $\TTb$-like flow \eqref{4dflow_gauge} with Maxwell's electromagnetism as its seed theory. Nevertheless, the discovery of the extent to which the dressing mechanism can reach was, from our perspective, quite surprising.
\subsubsection*{Massive electrodynamics and higher-order self-interactions}
Maxwell's theory \eqref{max_action} is the simplest theory of massless vector boson in four dimensions. At the price of breaking gauge invariance, one can add extra terms to the action, including a Lorentz-invariant potential $V$:
\begin{equation}\label{proca_action}
    S[g_{\mu\nu},A_\mu] = \int\dd^4 \Bx \sqrt{-g}\left[-\frac{1}{4}F_{\mu\nu}F^{\mu\nu} -V(A_\mu A^\mu)\right]\,.
\end{equation}
As an example, the choice $V =\frac{1}{2} m^2 A_\mu A^\mu$ reproduces the Proca action, describing a massive spin-1 vector field in $d=4$ space-time. The addition of an extra potential term to the Maxwell action has the simple effect of shifting the stress-energy tensor by a field-dependent quantity: 
\begin{equation}\label{shift}
    \BT_0 \to \BT_0 - V\mathbf{1}\,.
\end{equation}
In turn, equation \eqref{shift} produces a homogeneous shift in the eigenvalues of $\BT_0$, which does not however spoil the degeneracy discussed in \eqref{degeneracy_max}. This implies that the dressing procedure remains a feasible approach when performing $\TTb$-like deformations of self-interacting theories of electrodynamics. In particular, one can check that the interacting theory \eqref{proca_action} is deformed into \cite{Conti_Iannella}
\begin{equation}\label{contiiannella}
    S_\ta[g_{\mu\nu},A_\mu] = \frac{1}{2\ta\left(1-\ta V\right)}\int\dd^4\Bx\sqrt{-g}\left(\sqrt{\det\left[\mathbf{1}+\sqrt{2 \tau\left(1-\ta V\right)}\BF\right]}-1\right) + \frac{V}{1-\ta V}\,.
\end{equation}
It is interesting to observe that the effects of adding interaction terms to \eqref{max_action} amounts to a shift in the flow parameter in the kinetic term of the form
\begin{equation}
    \ta \rightarrow \ta \left(1-\ta V\right)\,.
\end{equation}
This property is not confined to Maxwell's theory. Rather, it extends its reach to generic Abelian gauge theories.

Beyond breaking gauge invariance, additional problems arise when introducing actions such as \eqref{proca_action}. These complications are ultimately related to discontinuous changes in the number of degrees of freedom in the theory when the potential vanishes. The Stueckelberg mechanism \cite{Stueckelberg:1957zz} could offer an efficient way to overcome these issues.

\subsubsection*{Deforming arbitrary Abelian gauge theories}

So far, we only focused on Maxwell's theory. It turns out that the class of physical theories which satisfy the bound provided by equation \eqref{eq:Ddiag} is much broader, encompassing all four-dimensional Abelian gauge theories. To show this, we turn our attention towards gauge theories built from the electromagnetic stress-tensor $F_{\mu\nu}$, relying on the results from \cite{Ferko_2022}. According to the Cayley-Hamilton theorem, every matrix $\BF$ satisfies its characteristic equation:
\begin{equation}\label{characteristic}
    \det\left[\BF\right] \mathbf{1} = \sum_{n=0}^3 \gamma_n \BF^n.
\end{equation}
The coefficients $\gamma_n$ are given by
\begin{equation}
    \gamma_n=\sum_{\left\{k_i\right\}} \prod_{i=1}^4 \frac{(-1)^{k_i+1}}{i^{k_i} k_{i} !}\tr\left[\BF^i\right]^{k_i},
\end{equation}
where the sum runs over all sets of non-negative integers $k_i$ which satisfy
\begin{equation}
    \sum_{i=1}^4 i k_i=3-n\,.
\end{equation}
Notice that, as the coefficients $\gamma_n$ depend only on the traces $\tr\left[\BF^i\right]$ for $i=1,\dots,4$, equation \eqref{characteristic} imposes constraints on the number of independent trace structures that a $4\times 4$ matrix may possess. Specifically, for $l>4$, any term proportional to $\tr\left[\BF^l\right]$ can be expressed in terms of lower traces. Furthermore, owing to the antisymmetry of the Abelian field strength $F_{\mu\nu}$, the trace of any odd power of its associated matrix $\BF$ is zero. Consequently, any scalar, gauge-invariant quantity constructed from $F_{\mu\nu}$ can be represented as a linear combination of the two remaining independent traces, namely $\tr\left[\BF^2\right]$ and $\tr\left[\BF^4\right]$. In particular, given that the Lagrangian densities $\cL_{U(1)}$ of four-dimensional Abelian gauge theories are by definition gauge-invariant scalars, one has, in full generality:
\begin{equation}\label{u1lagrangians}
    \cL_{U(1)} \left(g_{\mu\nu}, A_{\mu}\right)=  \cL_{U(1)} \left(\tr\left[\BF^2\right], \tr\left[\BF^4\right]\right).
\end{equation}
The Hilbert stress-energy tensor associated with the family of Lagrangians \eqref{u1lagrangians} is
\begin{equation}\label{tensoru(1)}
\begin{split}
     T_{\ta_0,\mu \nu}&=  g_{\mu \nu}\mathcal{L}_{U(1)}-2\sum_{n=1,\,2} \frac{\partial \mathcal{L}_{U(1)}}{\partial \tr\left[\BF^{2n}\right]} \cdot \frac{\delta \tr\left[\BF^{2n}\right]}{\delta g^{\mu \nu}}\\
     &=g_{\mu \nu}\mathcal{L}_{U(1)}-2\,\frac{\partial \mathcal{L}_{U(1)}}{\partial \tr\left[\BF^{2}\right]} \cdot \frac{\delta \tr\left[\BF^{2}\right]}{\delta g^{\mu \nu}} - 4\,\frac{\partial \mathcal{L}_{U(1)}}{\partial \tr\left[\BF^{4}\right]} \cdot \frac{\delta \tr\left[\BF^{4}\right]}{\delta g^{\mu \nu}}\;.
\end{split}
\end{equation}
Equation \eqref{tensoru(1)} can be written in terms of the associated matrices as
\begin{equation}
    \BT_{\ta_0}  = a_0 \mathbf{1} + a_1 \BF^2 + a_2\BF^4,
\end{equation}
where $a_i,\, i=0,\dots,2$, are some functions of $\tr\left[\BF^{2}\right]$ and $\tr\left[\BF^{4}\right]$.  If we label as $\{\ell_n\}_{n\in\{0,\dots,3\}}$ the eigenvalues of $\BF$, the eigenvalues of $\BT_{\ta_0}$ will result into 
\begin{equation}
    \{\lambda_n\}_{n\in\{0,\dots,3\}} =  \{a_0+a_1\ell^2_n + a_2\ell^4_n\}_{n\in\{0,\dots,3\}}\,.
\end{equation}
Similarly as with Maxwell's theory, since the eigenvalues of any antisymmetric matrix are purely imaginary and come in complex conjugate pairs, we take
\begin{equation}
\left\{\ell_0,\ell_1,\ell_2,\ell_3\right\} = \left\{-\ell_1,\ell_1,\ell_2,-\ell_2\right\}.
\end{equation}
Given that only even powers of each $\ell_n$ contributes to the set of eigenvalues of $\BT_{\ta_0}$, the latter can be put in a diagonal form $\BD$ by some similarity transformation, with
\begin{equation}\label{degeneracy_max_2}
  \BD = \text{diag}\left(\lambda_1,\lambda_1,\lambda_2,\lambda_2\right)\,.
\end{equation}
This tells us that, given any Abelian gauge theory in $d=4$ space-time, performing a $\TTb$-like deformation driven by the operator
\begin{equation}
   \mathcal{O}_{4,\ta}^{[1/2,1]} = \frac{1}{4}\left(\frac{1}{2}\tr\left[\BT_\ta\right]^2-\tr\left[\BT^2_\ta\right]\right)   
\end{equation}
is equivalent to implementing the dressing mechanism 
\begin{equation} \label{mnbvc}
    \left.  \left\{S_{U(1)}[g_{\mu\nu},A_\mu] -{\frac{\left(\ta-\ta_0 \right)}{4}} \int \mathrm{d}^4\Bx\sqrt{-g} \left(\frac{1}{2}\tr\left[\BT_{\ta_0}\right]^2-\,\tr\left[\BT^2_{\ta_0}\right]\right)\right\}\right|_{g=g(h)}  = S_{U(1),\ta}\left[h_{\var,\mu\nu}, A_\mu\right]\,,
\end{equation}
with deformed metric
\begin{equation}\label{h4dgauge}
    h_{\ta,\mu\nu} = 
    g_{\mu\nu} +\left(\var-\var_0\right)\left[T_{\var_0,\mu\nu} - \frac{1}{2}\tr\left[\BT_{\var_0}\right]g_{\mu\nu}\right].
\end{equation}
Given that gauge symmetry remains unbroken along $\mathrm{T}\overline{\mathrm{T}}$-like flows \cite{Ferko:2023wyi}, the characteristic degeneracy of the stress-energy tensor is preserved as we deform the theory. Then we see from \eqref{determinant_form_u1} that if the matter content of a four-dimensional $U(1)$ gauge theory $S_{U(1),\ta}$ complies with equation \eqref{mnbvc}, it must also satisfy
\begin{equation}\label{need_this_later}
   \left.  \left\{S_{U(1)}[g_{\mu\nu},A_\mu] -{{\left(\ta-\ta_0 \right)}} \int \mathrm{d}^4\Bx\sqrt{-g} \sqrt{\det\left[\BT^2_{\ta_0}\right]}\right\}\right|_{g=g(h)}  = S_{U(1),\ta}\left[h_{\var,\mu\nu}, A_\mu\right]\,.   
\end{equation}
As an example of non-standard electrodynamics subject to $\TTb$-like flows in four dimensions, consider the ModMax Lagrangian \cite{Bandos:2020jsw}, introduced as the unique one-parameter family of Lorentz invariant modifications
of Maxwell's theory preserving both duality invariance and conformal
symmetry:
\begin{equation}\label{modmax}
 \cL_{\mathrm{ModMax}}(g_{\mu\nu},A_\mu) = - \frac{1}{4}\cosh \gamma F_{\mu\nu}F^{\mu\nu}+ \frac{1}{4}\sinh \gamma \sqrt{\left(F_{\mu\nu}F^{\mu\nu}\right)^2+(\epsilon^{\mu\nu\alpha\beta}F_{\mu\nu}F_{\alpha\beta})^2}  . 
\end{equation}
Here $\gamma$ is a dimensionless real parameter that controls the deformation: when $\gamma$ is set to zero, the
Lagrangian \eqref{modmax} reduces to the usual Maxwell theory. When dressed by the $\TTb$-like deformation, it is deformed into the Born-Infeld-ModMax theory \cite{Babaei-Aghbolagh:2022uij, Ferko:2022iru}.
\begin{equation}
 \cL_{\mathrm{ModMax},\ta}(g_{\mu\nu},A_\mu)=\frac{1}{2\ta}\left[\sqrt{1+\ta\cL_\mathrm{ModMax}(g_{\mu\nu},A_\mu) -\frac{\ta^2}{4}(\epsilon^{\mu\nu\alpha\beta}F_{\mu\nu}F_{\alpha\beta})^2}-1\right]\,.
\end{equation}
As discussed before, gauge-breaking self-interactions will simply shift the eigenvalues of $\BT_{\ta_0}$, preserving their degeneracy, and the dressing procedure can be implemented also in the presence of non-vanishing potentials. As anticipated, the effects of the additional potential on the theory's kinetic term can be seen as a shift of the flow parameter $\ta \to \ta\left(1-\ta V\right)$. The interacting term is instead deformed as in \eqref{contiiannella}: $V \to V/\left({1-\ta V}\right)$.  This behaviour is a typical characteristic of $\TTb$-like Abelian gauge theories, and it reproduces the form of two-dimensional $\TTb$-deformed interacting theories \cite{Conti_Iannella}. 
\section{Palatini gravity}\label{section3}
In the past century, considerable exploration has been undertaken into extending general relativity in minimal ways, offering opportunities to depart from the traditional Einstein-Hilbert framework and tweak gravitational dynamics. Among these efforts, a particularly promising category is the $f(\cR)$ Palatini theories \cite{Olmo:2011uz} and their offshoots, the $f(\cR_{\mu\nu})$ extensions, also dubbed as Ricci-based gravity theories \cite{Olmo_2022}, which do not require the introduction of additional degrees of freedom compared to standard Einstein-Hilbert General Relativity. Before plunging into the subtleties of these theories, let us briefly review how in practice the Palatini formalism works.
For this purpose, we focus on the Einstein-Hilbert action:\footnote{In this paper, we set the reduced Planck mass $8\pi G_\mathrm{N} =1$.}\footnote{While the dimension $d$ remains arbitrary, caution is warranted in specific instances. Notably, when $d=2$ or $d=3$, the theory becomes topological, precluding the possibility of locally coupling matter degrees of freedom to gravity.} 
\begin{equation}\label{eh_palatini}
    S_{\mathrm{EH}} \left[g_{\mu\nu}, \Gamma^{\lambda}_{\mu\nu}\right] = \frac{1}{2}\int \dd^d\Bx\sqrt{-g}\, \tr\left[\BR(\Gamma)\right]\,.
\end{equation}
Notice that the action \eqref{eh_palatini} has a functional dependence from the connection $\Gamma$: this tacitly denotes our commitment to the Palatini framework. Herein, both the metric and the connection are regarded as independent dynamical fields, and the Ricci curvature tensor 
\begin{equation}
    \mathcal{R}_{\mu \nu} = \partial_\alpha \Gamma_{\nu \mu}^\alpha-\partial_\nu \Gamma_{\alpha \mu}^\alpha+\Gamma_{\alpha \beta}^\alpha \Gamma_{\nu \mu}^\beta-\Gamma_{\nu \beta}^\alpha \Gamma_{\alpha \mu}^\beta
\end{equation}
is considered as a functional of the connection only. Here 
\begin{equation}
\BR=\left(g^{\mu\alpha}\cR_{(\alpha\nu)}\right)_{\mu, \nu \in\{0, \ldots, d-1\}}     
\end{equation}
is the $d\times d$ matrix canonically associated with the symmetric part of the Ricci curvature tensor $\cR_{(\mu\nu)}$, where indices are as usual raised and lowered through the action of the metric tensor. The fact that only the symmetric part of $\cR_{\mu\nu}$ enters the action might seem inconsequential at present, but its significance will unveil in due course, as it will guarantee the stability of more complicated theories. Being $g_{\mu\nu}$ and $\Gamma^\lambda_{\mu\nu}$ two independent objects, the equations of motion for the theory \eqref{eh_palatini} are obtained by extremising the action with respect to both the metric and the connection. The variation of $S_{\mathrm{EH}}$ with respect to $g_{\mu\nu}$ yields
\begin{equation}\label{eh-eqs}
    \cR_{(\mu\nu)}(\Gamma)-\frac{1}{2}\tr[\BR(\Gamma)]g_{\mu\nu} = 0\,,
\end{equation}
while -- disregarding boundary contributions -- the variation with respect to $\Gamma^\lambda_{\mu\nu}$ gives us
\begin{equation}\label{conn-eqs}
 \Gamma^\lambda_{\mu\nu} =\frac12 g^{\lambda\alpha} \left(\partial_\nu g_{\mu\alpha} + \partial_\mu g_{\alpha\nu} - \partial_\alpha g_{\mu\nu}\right) \,.  
\end{equation}
Equations \eqref{eh-eqs} are nothing more than Einstein's vacuum field equations, albeit with the Ricci curvature depending upon the Palatini connection. But there is a catch: equation \eqref{conn-eqs} tells us that, when on-shell, $\Gamma^\lambda_{\mu\nu}$ is the usual Levi-Civita connection from General Relativity. At least when in vacuum, the dynamics of the Palatini-Einstein-Hilbert action \eqref{eh_palatini} yields the same as the one described by General Relativity. In other words, they are dynamically equivalent. What happens when we add matter to our theory? For our purpose, it is handy to consider a matter theory $S_\mathrm{M} = S_{\mathrm{M},\ta_0}$ as being the seed theory associated with some $\TTb$-like flow, as the ones described in the previous sections. Assuming that matter minimally couples with gravity, we write down the full theory as
\begin{equation}\label{eq:grav+matter}
    S \left[g_{\mu\nu}, \Gamma^\lambda_{\mu\nu} ,\Phi_I\right]=   S_{\mathrm{EH}}\left[g_{\mu\nu}, \Gamma^\lambda_{\mu\nu}\right] +  S_\mathrm{M}\left[g_{\mu\nu}, \Phi_I\right]\,.
\end{equation}
It is important to underline that in \eqref{eq:grav+matter} we assumed that the connection $\Gamma^\lambda_{\mu\nu}$ does not explicitly enter the matter action $S_\mathrm{M}$. This statement is equivalent to the condition that the theory must have vanishing hypermomentum:
\begin{equation}\label{hyper}
    \Delta^{\mu\nu}_\lambda = \frac{-2}{\sqrt{-g}}\frac{\delta S_\mathrm{M}}{\delta \Gamma^\lambda_{\mu\nu}} = 0\,.
\end{equation}
In other words, we are stating that matter fields refrain from direct coupling with the connection. This scenario generally holds for minimally coupled bosonic and Abelian gauge fields. Nevertheless, complications may emerge in the context of fermionic matter. For the sake of clarity, unless explicitly stated otherwise, we shall assume that all the matter theories we consider adhere to the condition \eqref{hyper}. As for the equations of motion of \eqref{eq:grav+matter}, notice that, since $\Gamma^\lambda_{\mu\nu}$ plays no role in the matter sector, equation \eqref{conn-eqs} is still valid: the connection for our theory is once again the Levi-Civita one. On the other hand, since $g_{\mu\nu}$ takes part in $S_\mathrm{M}$, it is easy to see that the equations associated with the metric tensor become
\begin{equation}\label{eh-eq-T}
 \cR_{(\mu\nu)}-\frac{1}{2}\tr[\BR]g_{\mu\nu} = T_{\ta_0,\mu\nu}\,,  
\end{equation}
where we finally dropped the explicit $\Gamma^\lambda_{\mu\nu}$ dependence, as a way of indicating that the connection is the metric-compatible one. Equations \eqref{eh-eq-T} are, once again, equivalent to those from General Relativity. However, the genuine strength of the Palatini approach comes to light when we depart from the conventional Einstein-Hilbert action, as we will see shortly.
\subsection{$f(\cR)$ and Ricci-based gravity theories}
One of the most straightforward ways to increase the complexity of the gravity functional is to add higher-order terms to the action \eqref{eh_palatini}. For example, one may decide to consider sufficiently well-behaved functionals, which can be then Taylor expanded into
\begin{equation}\label{general_grav}
 S_{\mathrm{G},\vargrav}\left[g_{\mu\nu}, \Gamma^\lambda_{\mu\nu}\right]  = \int \dd^d \Bx \sqrt{-g}\sum_{m,n}\,c_{mn}\,\vargrav^{n+m-1}\tr[\BR(\Gamma)]^m\tr[\BR^n(\Gamma)] \,.
\end{equation}
Note that when writing down the action \eqref{general_grav} we were somehow forced to incorporate at least one-dimensional parameter, which we denoted as $\vargrav$, with the convention that $[\vargrav]=-2$ in mass units, so to ensure that $[S_{\mathrm{G},\vargrav}]=1$. The presence of $\vargrav$ sets a mass scale for the theory. Here the $c_{mn}$'s are coefficients which define the theory. While refraining from detailing the specific structure of the action \eqref{general_grav}, it is understandable to require that its IR limit should converge to the Einstein-Hilbert action:
\begin{equation}\label{norm}
 S_{\mathrm{G},\vargrav}\left[g_{\mu\nu}, \Gamma^\lambda_{\mu\nu}\right] = S_{\mathrm{EH}}\left[g_{\mu\nu}, \Gamma^\lambda_{\mu\nu}\right] + O(\vargrav)\,.  
\end{equation}
This is easily achieved by setting $c_{00}=0$\footnote{The $c_{00}$ term simply amounts to the addition of a cosmological constant to the model, contributing to its vacuum energy. While our present discussion can be easily generalised to the case of non-vanishing cosmological constant, we will be ultimately interested in the analysis of asymptotically flat space-time.} and $c_{10}=c_{01}=1/4$ (notice that these last two coefficients are associated to identical terms in \eqref{general_grav}). When $c_{mn} = 0$ for every $n\neq 0$, the resulting theory can be Taylor expanded into powers of $\tr[\BR]$ alone, and things get significantly simpler. These theories are known as $f(\mathcal{R})$ theories, and we will explore them in the following section. When there are no constraints on the coefficients $c_{mn}$ -- apart from those realising the low energy limit \eqref{norm} -- the theories in the class parametrised by \eqref{general_grav} are generally known as Ricci-based gravity theories (RBGs for short). It is within the context of RBGs that our restriction to the symmetric component of $\cR_{\mu\nu}$ becomes relevant: as shown in \cite{BeltranJimenez:2019acz}, even though the field equations of RBGs are, in general, of second order, it is necessary to impose projective invariance of the action to get rid of ghost-like dynamical degrees of freedom. A projective transformation of the connection field of the form
\begin{equation}\label{projinv}
\Gamma_{\mu \nu}^\lambda\to\Gamma_{\mu \nu}^\lambda+\xi_\mu \delta_\nu^\lambda\,,
\end{equation}
reflects on the Ricci curvature tensor as
\begin{equation}
    \cR_{\mu\nu} \to \cR_{\mu\nu} + \partial_{\mu}\xi_\nu - \partial_\nu\xi_\mu\,.
\end{equation}
Consequently, the symmetric part of $\cR_{\mu\nu}$ remains unaffected, and so does the gravitational action $S_{\mathrm{G},\vargrav}$.
\subsubsection{$f(\cR)$ gravity}
In this section, setting $c_{mn} = 0$ for every $n\neq 0$ in \eqref{general_grav}, we focus on the family of functionals
\begin{equation}\label{general_grav_fr}
 S_{\mathrm{G},\vargrav}\left[g_{\mu\nu}, \Gamma^\lambda_{\mu\nu}\right]  = \int \dd^d \Bx \sqrt{-g}\left(\frac{1}{2}\tr[\BR(\Gamma)] + \sum_{m\geq2}c_{m0} \vargrav^{m-1}\tr[\BR(\Gamma)]^m\right)\,.
\end{equation}
It turns out that one of the most notable members of this class is also the simplest, non-trivial example of Palatini $f(\cR)$ gravity in $d=4$, known as the Starobinsky model:
\begin{equation}\label{Starobinsky-4-d}   S_{\mathrm{Star},\vargrav}\left[g_{\mu\nu},\Gamma^\lambda_{\mu\nu}\right] = \int \mathrm{d}^4\Bx\,\sqrt{-g} \left(\frac{1}{2}\tr\left[\BR(\Gamma)\right] + \frac{\vargrav}{4}\tr\left[\BR(\Gamma)\right]^2\right)\,.
\end{equation}
The action \eqref{Starobinsky-4-d} will later play a central role in this article, but we will keep the discussion as general as possible for the moment being. As we did for the Einstein-Hilbert action in the Palatini framework, we add matter sources, minimally coupling the gravity theory to an arbitrary seed theory in $\ta=\ta_0$:
\begin{equation}\label{total_f(r)}
   S_\vargrav\left[g_{\mu\nu}, \Gamma^\lambda_{\mu\nu} ,\Phi_I\right]=   S_{\mathrm{G},\vargrav}\left[g_{\mu\nu}, \Gamma^\lambda_{\mu\nu}\right] +  S_\mathrm{M}\left[g_{\mu\nu}, \Phi_I\right]\,.   
\end{equation}
For the sake of notational compactness, from now on we will drop the explicit $\Gamma$ dependency in $\cR_{\mu\nu}$. It is then convenient to introduce a gravitational Lagrangian density $\cL_{\mathrm{G},\vargrav}$ such that $S_{\mathrm{G},\vargrav} = \int \mathrm{d}^4\Bx\,\sqrt{-g}\,\cL_{\mathrm{G},\vargrav}$. Since, by assumption, $\cL_{\mathrm{G},\vargrav}$ should exclusively depend on $g_{\mu\nu}$ through the trace of the matrix $\BR$, we can trade the metric variation of the gravitational Lagrangian with its variation with respect to $\tr[\BR]$:
\begin{equation}
 \frac{\partial \cL_{\mathrm{G},\vargrav}}{\partial g^{\mu\nu}} =  \frac{\partial \cL_{\mathrm{G},\vargrav}}{\partial \tr[\BR]}\frac{\partial \tr[\BR]}{\partial g^{\mu\nu}}  =  \frac{\partial \cL_{\mathrm{G},\vargrav}}{\partial \tr[\BR]} \cR_{(\mu\nu)}\,.
\end{equation}
Requiring that the variation of $S_{\mathrm{G},\vargrav}$ with respect to $g_{\mu\nu}$ should vanish, we obtain
\begin{equation}\label{eoms_for_f(r)}
\begin{split}
2\frac{\partial \cL_{\mathrm{G},\vargrav}}{\partial \tr[\BR]}\cR_{(\mu\nu)} - \cL_{\mathrm{G},\vargrav} g_{\mu\nu} =T_{\ta_0,\mu\nu}\,.    
\end{split}
\end{equation}
Finally, taking the trace of equation \eqref{eoms_for_f(r)}, we get
\begin{equation}\label{traceofeoms}
 2\frac{\partial \cL_{\mathrm{G},\vargrav}}{\partial \tr[\BR]}\tr[\BR] - d\cL_{\mathrm{G},\vargrav} =\tr[\BT_{\ta_0}]  \,. 
\end{equation}
This equation implies that $\tr[\BR(\Gamma)]$ can be solved algebraically in terms
of $\tr[\BT_{\ta_0}]$, thus leading to $\tr[\BR(\Gamma)] = f(\tr[\BT_{\ta_0}])$, as a function of the matter content (and possibly of the metric), but not of the independent connection. On the other hand, using
\begin{equation}
    \delta \mathcal{R}_{(\mu \nu)}={\nabla}_\lambda \delta \Gamma_{\mu \nu}^\lambda-{\nabla}_\nu \delta \Gamma_{\mu \lambda}^\lambda\,,
\end{equation}
where $\nabla_\mu$ denotes the covariant derivative associated to $\Gamma_{\mu \nu}^\lambda$, the variation of \eqref{total_f(r)} with respect to the connection yields
\begin{equation}\label{eoms_conn_f(r)}
   {\nabla}_\lambda\left(\sqrt{-g} \frac{\partial \cL_{\mathrm{G},\vargrav}}{\partial \tr[\BR]}g^{\mu \nu}\right)=0 .
\end{equation}
We can introduce an auxiliary metric tensor $h_{\vargrav,\mu\nu}$, defined by its inverse $\left(h^{-1}_\vargrav\right)^{\mu\nu}$ via
\begin{equation}\label{mcomp}
  \frac{1}{2}\sqrt{-h_\vargrav}\left(h^{-1}_\vargrav\right)^{\mu\nu} = \sqrt{-g}\frac{\partial \cL_{\mathrm{G},\vargrav}}{\partial \tr[\BR]}g^{\mu\nu}\,,\qquad h_\vargrav := \det\left[h_{\vargrav,\mu\nu}\right],
\end{equation}
with $\left(h^{-1}_\vargrav\right)^{\mu\alpha}h_{\vargrav,\alpha\nu} = \delta^\mu_\nu$. Notice that the new metric $h_{\vargrav,\mu\nu}$ is symmetric by construction, and it converges to the standard metric $g_{\mu\nu}$ for sufficiently small values of the coupling $\vargrav$, as one can see from \eqref{general_grav_fr}. The additional $1/2$ factor in \eqref{mcomp} may appear arbitrary at first glance, yet its significance will become apparent in equation \eqref{sigmaos}.

With this definition in mind, we can recast the equations of motion for the connection \eqref{eoms_conn_f(r)} into a far more familiar form:
\begin{equation}\label{compatible}
\Gamma^\lambda_{\mu\nu} =\frac12 \left(h_\vargrav^{-1}\right)^{\lambda\alpha} \left(\partial_\nu h_{\vargrav,\mu\alpha} + \partial_\mu h_{\vargrav,\alpha\nu} - \partial_\alpha h_{\vargrav,\mu\nu}\right)\,.   
\end{equation}
Equation \eqref{compatible} simply states that the auxiliary metric $h_{\vargrav,\mu\nu}$ is such that the connection must be $h_\vargrav$-compatible. Moreover, taking the determinant of both sides of \eqref{mcomp}, and plugging back in the value of $\sqrt{-h_\vargrav}$, equation \eqref{mcomp} can be simplified into
\begin{equation}\label{h_explicit}
    h_{\vargrav,\mu\nu} = \left(2\frac{\partial \cL_{\mathrm{G},\vargrav}}{\partial \tr[\BR]}\right)^{\frac{2}{d-2}}g_{\mu\nu}\,.
\end{equation}
This tells us that $h_{\vargrav,\mu\nu}$ and $g_{\mu\nu}$ are related by a theory-dependent Weyl rescaling. Relying on the fact that equation \eqref{traceofeoms} allows us to trade traces of $\BR$ with functionals of the trace of the stress-energy tensor $\BT_{\ta_0}$, the auxiliary metric and the standard one are ultimately linked by the matter content of the full theory. Notice that the dimension of space-time plays an important role in defining the behaviour of the auxiliary metric.

An interesting question arises: what happens in vacuum, when $\BT_{\ta_0}=0$? From the equations of motion \eqref{traceofeoms}, we notice that $\tr[\BR]$ is fixed as a constant as we move throughout spacetime, and so is $\cL_{\mathrm{G},\vargrav}$. Introducing the auxiliary metric $h_{\vargrav,\mu\nu}$ simply amounts to some homogeneous rescaling of $g_{\mu\nu}$, and the connection becomes $g$-compatible. In the absence of sources, the dynamics of Palatini $f(\cR)$ gravity simplifies to that of General Relativity, albeit with the potential addition of an effective cosmological constant to accommodate the correct fixed value for the scalar curvature $\tr[\BR]$ \cite{Olmo:2011uz}.

\subsubsection{Ricci-based gravity}
Despite their increased complexity, generic RBGs share various features with $f(\cR)$ theories, and the techniques employed for their study are essentially the same. This time, since $\cL_{\mathrm{G},\vargrav}$ will depend on some functional of the matrix $\BR$, we can trade the metric variation of the gravitational Lagrangian with its variation with respect to the combination $g^{\mu\alpha} \cR_{(\alpha\nu)}$, since
\begin{equation}\label{maeoms2}
    \frac{\partial \cL_{\mathrm{G},\vargrav}}{\partial g^{\mu\nu}} =  \frac{\partial \cL_{\mathrm{G},\vargrav}}{\partial g^{\rho\alpha}\cR_{(\alpha\beta)}}\frac{\partial g^{\rho\sigma}\cR_{(\sigma\beta)}}{\partial g^{\mu\nu}} = \frac{\partial \cL_{\mathrm{G},\vargrav}}{\partial g^{\mu\alpha}\cR_{(\alpha\beta)}} \cR_{(\beta\nu)}\,.
\end{equation}
Once the gravity theory has been suitably coupled with matter, the variation of the total action with respect to the metric yields
\begin{equation}\label{eoms_RBG}
   2\frac{\partial \cL_{\mathrm{G},\vargrav}}{\partial g^{\mu\alpha}\cR_{(\alpha\beta)}} \cR_{(\beta\nu)} - \cL_{\mathrm{G},\vargrav}g_{\mu\nu} = T_{\ta_0,\mu\nu}\,.
\end{equation}
As far as the equations of motion for the independent connection are concerned, we can once again introduce an auxiliary metric tensor $h_{\vargrav,\mu\nu}$ such that
\begin{equation}\label{compatibility_RBGs}
 \Gamma^\lambda_{\mu\nu} =\frac12 \left(h_\vargrav^{-1}\right)^{\lambda\alpha} \left(\partial_\nu h_{\vargrav,\mu\alpha} + \partial_\mu h_{\vargrav,\alpha\nu} - \partial_\alpha h_{\vargrav,\mu\nu}\right)\,.    
\end{equation}
In this case, the compatibility condition \eqref{compatibility_RBGs} is realised by defining
\begin{equation}\label{hforrbg}
      \frac{1}{2}\sqrt{-h_\vargrav}\left(h_\vargrav^{-1}\right)^{\mu\nu} = \sqrt{-g}g^{\mu\alpha}\frac{\partial \cL_{\mathrm{G},\vargrav}}{\partial g^{\alpha\beta}\cR_{(\beta\nu)}}\,.
\end{equation}
Just as for the $f(\cR)$ case, in vacuum, the field equations of RBGs \eqref{eoms_RBG} coincide with those of General Relativity \cite{Ferraris:1992dx}. This is a common feature of metric-affine gravity theories, where modifications to the gravitational sector emerge only in the presence of matter fields. The universality of this outcome ultimately stems from the covariance of the field equations, which implies that -- in a vacuum -- the Ricci tensor must be proportional to the metric. An additional perspective on this broad result arises from recognising that General Relativity is the unique Lorentz invariant and unitary theory governing a self-interacting, massless spin-$2$ field, commonly referred to as the graviton. Thus, if by gravity we mean a theory for such a particle, we will inevitably find General Relativity in vacuum \cite{jimenez_2018}.
\subsection{The two frames of Palatini gravity}
It has been shown \cite{Magnano_1990, Delsate:2013bt, Olmo_2022} that Ricci-based gravity theories such as \eqref{eq:grav+matter} can be recast into General Relativity, provided we introduce modified couplings in the matter sector of the full action. To see how this works, we start by focusing on the $f(\cR)$ case, and introduce the auxiliary action
\begin{equation}\label{aux}
\begin{split}
\widehat{S}_{\mathrm{G},\vargrav}\left[g_{\mu\nu}, \Gamma^\lambda_{\mu\nu}, \aux^\mu_\nu\right] &= \int \mathrm{d}^d \Bx \sqrt{-g} \left\{\cL_{\mathrm{G},\vargrav}\left(\tr[\bm{\aux}]\right)+\left(\tr[\BR]-\tr[\bm{\aux}]\right)\frac{\partial \cL_{\mathrm{G},\vargrav}}{\partial \tr[\bm{\aux}]}\right\},
\end{split}
\end{equation}
and we couple it to the same matter content introduced in the previous section:
\begin{equation}\label{aux+matter}
    \widehat{S}_\vargrav \left[g_{\mu\nu}, \Gamma^\lambda_{\mu\nu}, \aux^\mu_\nu, \Phi_I\right] = \widehat{S}_{\mathrm{G},\vargrav}\left[g_{\mu\nu}, \Gamma^\lambda_{\mu\nu}, \aux^\mu_\nu\right] + S_\mathrm{M}\left[g_{\mu\nu}, \Phi_I\right]\,.
\end{equation}
Importantly, this theory depends on the additional matrix field $\bm{\aux}$, which will prove crucial in establishing a dynamical equivalence between $S_\vargrav$ and the auxiliary action $\widehat{S}_\vargrav$. For $f(\cR)$ theories, the additional dependence of $\widehat{S}_\vargrav$ is entirely summarised within the trace of $\bm{\aux}$, which can be regarded as a single auxiliary scalar field. Performing the variation of \eqref{aux+matter} with respect to $\tr[\bm{\aux}]$, we obtain
\begin{equation}\label{sigma_var}
 \frac{\delta\widehat{S}_\vargrav}{\delta\tr[\bm{\aux}]} = \int \mathrm{d}^d \Bx \sqrt{-g} \left\{\frac{\partial\cL_{\mathrm{G},\vargrav}}{\partial \tr[\bm{\aux}]}+\left(\tr[\BR] - \tr[\bm{\aux}]\right)\frac{\partial^2 \cL_{\mathrm{G}}}{\partial \tr[\bm{\aux}]^2}- \frac{\partial\cL_{\mathrm{G},\vargrav}}{\partial \tr[\bm{\aux}]}\right\},
\end{equation}
which, when set to zero, constrains the on-shell value of $\bm{\aux}$ to
\begin{equation}\label{on-shell-sigma}
\tr[\bm{\aux}^\star] = \tr[\BR]\;, 
\end{equation}
provided that the second derivative of $\cL_{\mathrm{G},\vargrav}$ in \eqref{sigma_var} does not vanish. Plugging \eqref{on-shell-sigma} back into \eqref{aux+matter}, we observe that,
\begin{equation}\label{equiv}
    \widehat{S}_\vargrav\left[g_{\mu\nu}, \Gamma^\lambda_{\mu\nu}, \left(\aux^\star\right)^\mu_\nu,\Phi_I\right] \simeq {S}_\vargrav\left[g_{\mu\nu}, \Gamma^\lambda_{\mu\nu}, \Phi_I\right],
\end{equation}
i.e., the two actions satisfy the same Euler-Lagrange equations. We notice moreover that the gravitational sector of $\widehat{S}_\vargrav$ can be recast into a more familiar structure by rearranging its terms as follows:
\begin{equation}\label{rhs}
\begin{split}
\widehat{S}_{\mathrm{G},\vargrav} &= \int \mathrm{d}^d \Bx \sqrt{-g}\,\tr[\BR]\frac{\partial \cL_{\mathrm{G},\vargrav}}{\partial \tr[\bm{\aux}]} + \int \mathrm{d}^d \Bx \sqrt{-g}\,\left(\cL_{\mathrm{G},\vargrav}- \tr[\bm{\aux}]\frac{\partial \cL_{\mathrm{G},\vargrav}}{\partial \tr[\bm{\aux}]}\right).
\end{split}
\end{equation}
At this point, we introduce yet another auxiliary metric tensor, which we suggestively call $h_{\vargrav,\mu\nu}$, defined by
\begin{equation}\label{sigmadef}
    \frac{1}{2}\sqrt{-h_\vargrav}\left(h_\vargrav^{-1}\right)^{\mu\nu}= \sqrt{-g}\,\frac{\partial \cL_{\mathrm{G},\vargrav}}{\partial \tr[\bm{\aux}]}g^{\mu\nu}\;.
\end{equation}
The resemblance to \eqref{mcomp} is not unintentional: once we integrate out the field $\bm{\aux}$, thus ensuring that $\widehat{S}_\vargrav\simeq {S}_\vargrav$, the value of $h_{\mu\nu}$ will be defined by
\begin{equation}\label{sigmaos}
   \frac{1}{2}\sqrt{-h_\vargrav}\left(h_\vargrav^{-1}\right)^{\mu\nu}= \sqrt{-g}\,\frac{\partial \cL_{\mathrm{G},\vargrav}}{\partial \tr[\BR]}g^{\mu\nu}\,.
\end{equation}
This quite elementary observation bears a profound meaning. The dynamical equivalence of $\widehat{S}_\vargrav$ and $S_\vargrav$ is manifested on configurations of $\bm{\aux}$ satisfying the equations of motion \eqref{on-shell-sigma}. Intriguingly, it so happens that these configurations are such that the auxiliary tensor $h_{\vargrav,\mu\nu}$ automatically meets the compatibility conditions for the connection. On the other hand, the introduction of $h_{\vargrav,\mu\nu}$ as a new metric allows isolating in equation \eqref{rhs} the standard Einstein-Hilbert action with respect to $h_{\vargrav,\mu\nu}$, and we can finally rewrite \eqref{aux+matter} as
\begin{equation}\label{rhs2}
\begin{split}
\widehat{S}_\vargrav &= \frac12\int \mathrm{d}^d \Bx \sqrt{-h_\vargrav}\,\tr_\vargrav[\BR] + \int \mathrm{d}^d \Bx \sqrt{-g}\,\left(\cL_{\mathrm{G},\vargrav}- \tr[\bm{\aux}]\frac{\partial \cL_{\mathrm{G},\vargrav}}{\partial \tr[\bm{\aux}]}\right)+ S_\mathrm{M}\left[g_{\mu\nu},\Phi_I\right]\\
&=\frac12\int \mathrm{d}^d \Bx \sqrt{-h_\vargrav}\,\tr_\vargrav[\BR] + {S}_{\mathrm{M},\ta_0 + \vargrav}\left[g_{\mu\nu}, \aux^\mu_\nu, \Phi_I\right]\,.
\end{split}
\end{equation}
Here we borrowed the notation introduced in section \ref{section2}, denoting by $\tr_\vargrav$ traces taken using the auxiliary metric $h_{\vargrav,\mu\nu}$. In the second line of \eqref{rhs2} we defined a new matter action ${S}_{\mathrm{M},\ta_0 + \vargrav}$, which incorporates the residual $\bm{\aux}$-dependence of the theory. In principle, the modified matter action can be rewritten in terms of $h_{\vargrav,\mu\nu}$ and $\Phi_I$ alone, by solving the definition \eqref{sigmadef} for $g_{\mu\nu}$.  This means that the dynamics of a matter theory $S_{\mathrm{M}}$ minimally coupled to $f(\cR)$ gravity is equivalent to that of a deformed matter theory $S_{\mathrm{M},\tau_0+\vargrav}$ minimally coupled to standard General Relativity. Here we say standard because the connection we are working with is, by definition of the auxiliary metric tensor, $h_\vargrav$-compatible.  Explicitly defining the stress-energy tensor
\begin{equation}\label{tildet}
    T_{\ta_0 + \vargrav,\mu\nu} = \frac{-2}{\sqrt{-h_\vargrav}}\frac{\delta {S}_{\mathrm{M},\ta_0 + \vargrav}}{\delta (h_\vargrav^{-1})^{\mu\nu}}\,,
\end{equation}
the equations of motion resulting from the variation of  \eqref{rhs2} with respect to $h_{\vargrav,\mu\nu}$ can be expressed as:
\begin{equation}\label{einst}
\cR_{(\mu\nu)} - \frac{1}{2}\tr_\vargrav[\BR]h_{\vargrav,\mu\nu} ={T}_{\ta_0 + \vargrav,\mu\nu}\,.
\end{equation}
As expected, the expression in \eqref{einst} corresponds to the Einstein field equations formulated in terms of the auxiliary metric tensor $h_{\mu\nu}$. These results can be generalised for arbitrary RBG theories, by introducing the auxiliary action
\begin{equation}\label{aux_rbg}
\begin{split}
\widehat{S}_{\mathrm{G},\vargrav}\left[g_{\mu\nu}, \Gamma^\lambda_{\mu\nu}, \aux^\mu_\nu\right] &= \int \mathrm{d}^d \Bx \sqrt{-g} \left\{\cL_{\mathrm{G},{\vargrav}}\left(\bm{\aux}\right)+\left(g^{\mu\alpha}\cR_{(\alpha\nu)} - \aux^\mu_\nu\right)\frac{\partial \cL_{\mathrm{G},\vargrav}}{\partial \aux^\mu_\nu}\right\}.
\end{split}
\end{equation}
Similarly, we have
 \begin{equation}\label{sigma_var_rbg}
 \frac{\delta\widehat{S}_\vargrav}{\delta  \aux^\alpha_\beta} = \int \mathrm{d}^d \Bx \sqrt{-g} \left\{\frac{\partial\cL_{\mathrm{G},\vargrav}}{\partial \aux^\alpha_\beta}+\left(g^{\mu\rho}\cR_{(\rho\nu)} - \aux^\mu_\nu\right)\frac{\partial^2 \cL_{\mathrm{G}}}{\partial \aux^\mu_\nu\partial \aux^\alpha_\beta}- \frac{\partial\cL_{\mathrm{G},\vargrav}}{\partial \aux^\alpha_\beta}\right\}\,,
\end{equation}
which forces $\bm{\aux}^\star = \BR$. When we substitute this result back into \eqref{aux_rbg}, we obtain the dynamical equivalence
\begin{equation}\label{equiv_rbg}
    \widehat{S}_\vargrav\left[g_{\mu\nu}, \Gamma^\lambda_{\mu\nu}, \left(\aux^\star\right)^\mu_\nu,\Phi_I\right] \simeq {S}_\vargrav\left[g_{\mu\nu}, \Gamma^\lambda_{\mu\nu}, \Phi_I\right]\,.
\end{equation}
Again, we carefully rearrange terms in \eqref{aux_rbg}, and write down the gravity sector as
\begin{equation}\label{shat_rbg}
 \begin{split}
\widehat{S}_{\mathrm{G},\vargrav} &= \int \mathrm{d}^d \Bx \sqrt{-g}\,g^{\mu\alpha}\cR_{(\alpha\nu)}\frac{\partial \cL_{\mathrm{G},\vargrav}}{\partial \aux^\mu_\nu} + \int \mathrm{d}^d \Bx \sqrt{-g}\,\left(\cL_{\mathrm{G},\vargrav}- \aux^\mu_\nu\frac{\partial \cL_{\mathrm{G},\vargrav}}{\partial \aux^\mu_\nu}\right).
\end{split}   
\end{equation}
As in \eqref{sigmadef}, we can introduce the auxiliary metric
\begin{equation}\label{sigmadef_rbg}
    \frac{1}{2}\sqrt{-h_\vargrav}\left(h_\vargrav^{-1}\right)^{\mu\nu}= \sqrt{-g}g^{\mu\alpha}\frac{\partial \cL_{\mathrm{G,\vargrav}}}{\partial H^\alpha_\nu}\;,
\end{equation}
which, to ensure that \eqref{equiv_rbg} is realised, assumes the familiar form
\begin{equation}\label{sigmaos_rbg}
   \frac{1}{2}\sqrt{-h_\vargrav}\left(h_\vargrav^{-1}\right)^{\mu\nu}= \sqrt{-g}g^{\mu\alpha}\frac{\partial \cL_{\mathrm{G},\vargrav}}{\partial g^{\alpha\beta}\cR_{(\beta\nu)}}\,.
\end{equation}
Noting that the first term in \eqref{shat_rbg} represents the Einstein-Hilbert action expressed in terms of the metric $h_{\vargrav,\mu\nu}$, we obtain
\begin{equation}
\begin{split}
\widehat{S}_\vargrav &= \frac12\int \mathrm{d}^d \Bx \sqrt{-h_\vargrav}\,\tr_\vargrav[\BR] + \int \mathrm{d}^d \Bx \sqrt{-g}\,\left(\cL_{\mathrm{G},\vargrav}- \aux^\mu_\nu\frac{\partial \cL_{\mathrm{G},\vargrav}}{\partial \aux^\mu_\nu}\right)+ S_\mathrm{M}\left[g_{\mu\nu},\Phi_I\right]\\
&=\frac12\int \mathrm{d}^d \Bx \sqrt{-h_\vargrav}\,\tr_\vargrav[\BR] + {S}_{\mathrm{M},\ta_0 + \vargrav}\left[g_{\mu\nu}, \aux^\mu_\nu, \Phi_I\right]\,.
\end{split}    
\end{equation}
This tells us that the equations of motion for $\widehat{S}_\vargrav$ can be written as
\begin{equation}\label{einst2}
\cR_{(\mu\nu)} - \frac{1}{2}\tr_\vargrav[\BR]h_{\vargrav,\mu\nu} ={T}_{\ta_0 + \vargrav,\mu\nu}\,,
\end{equation}
with ${T}_{\ta_0 + \vargrav,\mu\nu}$ defined as in \eqref{tildet}. We shall henceforth designate $\widehat{S}_\vargrav$ -- defined in terms of the new metric $h_{\vargrav,\mu\nu}$ and leading to the familiar Einstein equations -- as the \textit{Einstein frame} action, while denoting $S_\vargrav$ -- defined by the original metric $g_{\mu\nu}$ -- as the \textit{Palatini frame} action. In $f\left(\cR\right)$ theories, realised when the gravitational Lagrangian $\cL_{\mathrm{G},\ta}$ is a function of $\tr\left[\BR\right]$ only, the geometries of the two frames are conformally related \cite{Capozziello:2010ef}. There is a close analogy between the above discussion and the existence of two frames in scalar-tensor theories of gravity, which we address in section \ref{section4}. In the broader framework of RBGs, the connection equation can still be solved in terms of an auxiliary geometry which, nonetheless, is not conformal in general, but disformal \cite{Alinea:2020sei}. Recapitulating, the integration of $\bm{\aux}$ in $\widehat{S}_\vargrav$ automatically enforces the metric compatibility condition, while simultaneously ensuring $\widehat{S}_\vargrav\simeq S_\vargrav$. As a consequence, we can write the dynamical equivalence
\begin{equation}\label{equiv_between_frames}
S_{\mathrm{G},\vargrav} \left[g_{\mu\nu},\Gamma^\lambda_{\mu\nu}\right]+ S_{\mathrm{M}}\left[g_{\mu\nu},\Phi_I\right] \simeq S_{\mathrm{EH}}\left[h_{\vargrav,\mu\nu}\right] +   S_{\mathrm{M},\ta_0 + \vargrav}\left[h_{\vargrav,\mu\nu},\Phi_I\right]\,,
\end{equation}
together with the definition of the modified matter theory
\begin{equation}\label{corr-act}
\begin{split}
S_{\mathrm{M},\ta_0 + \vargrav}\left[h_{\vargrav,\mu\nu},\Phi_I\right] =\left.\left\{S_{\mathrm{M}}\left[g_{\mu\nu},\Phi_I\right]+\int \mathrm{d}^d\Bx\sqrt{-g}\,\left(\cL_{\mathrm{G},\vargrav}- g^{\mu\alpha}\cR_{(\alpha\nu)}\frac{\partial \cL_{\mathrm{G},\vargrav}}{\partial g^{\mu\beta}\cR_{(\beta\nu)}}\right)\right\}\right|_{g=g(h)}\!\! ,
\end{split}
\end{equation}
and of the auxiliary metric $h_{\vargrav,\mu\nu}$
\begin{equation}\label{defmetric_1}
     \frac{1}{2}\sqrt{-h_\vargrav}\left(h_\vargrav^{-1}\right)^{\mu\nu}= \sqrt{-g}g^{\mu\alpha}\frac{\partial \cL_{\mathrm{G},\vargrav}}{\partial g^{\alpha\beta}\cR_{(\beta\nu)}}\;.
\end{equation} 
Notice that we dropped the presence of $\Gamma^{\lambda}_{\mu\nu}$ in the right-hand side of \eqref{equiv_between_frames}, since it is realised as the Levi-Civita connection of the metric $h_{\vargrav,\mu\nu}$. More importantly, notice the resemblance between equation \eqref{corr-act} (on the gravity side) and equation \eqref{bare_deformation} (on the seemingly unrelated $\TTb$-like side). There are scenarios, as we shall soon discover, where these equations convey equivalent insights regarding the underlying physics.
\section{$\TTb$-like dressing from gravitational reframing}\label{section4}
As we noted before, there is an interesting similarity between equation \eqref{corr-act}, characterized by a metric deformation as defined by \eqref{defmetric_1}, and the $\TTb$-like dressing introduced in \eqref{bare_deformation}. The Legendre transform of $\cL_{\mathrm{G},\vargrav}$ appearing in \eqref{corr-act} can be expressed as a function of the Ricci tensor. If we then impose the field equations \eqref{eoms_RBG}, we can write \eqref{corr-act} in terms of the energy-momentum tensor $T_{\tau_0,\mu\nu}$ associated with the theory in the Palatini frame. Similarly, also the auxiliary metric $h_{\vargrav,\mu\nu}$ can be written as a function of the original metric $g_{\mu\nu}$ and the stress tensor of the matter sector. Then, if $\mathcal{L}_{\mathrm{G},\vargrav}$ is such that this rewriting produces the desired $\TTb$-like operator and the Einstein frame metric aligns with the corresponding $\TTb$-deformed metric, it automatically follows that the matter action in the Einstein frame is dynamically equivalent to the $\TTb$-like deformation of the Palatini frame matter action. As we will show through specific examples later on, our deliberate choice to have $S_{\mathrm{G},\vargrav}$ depended on the irrelevant parameter $\vargrav$ endows the latter with the role of a flow parameter $\tau$ governing the deformation, up to some linear transformation. We have fixed $[\vargrav]=-2$, so that standard dimensional analysis settles that the correspondence should be established for $\vargrav \propto \ta-c$, where $c$ is some constant. However, as our ultimate aim is to express \eqref{corr-act} in terms of the stress tensor of the Palatini frame theory, implementing consistent gluing conditions requires setting $c=\ta_0$, so that
\begin{equation}\label{choice_of_epsilon}
    \vargrav \propto \ta-\ta_0.
\end{equation}
In the end, the specific value of the proportionality constant in \eqref{choice_of_epsilon} turns out to be physically irrelevant, as we retain the flexibility to redefine the deforming parameter $\ta$ at will. Our discussion so far implies that appropriately crafted modified gravity theories can serve as a source for generating compatible $\TTb$-like deformations. Fixing $\vargrav$ as in \eqref{choice_of_epsilon}, and setting (without loss of generality) the proportionality constant in \eqref{choice_of_epsilon} to $1$, the dynamical equivalence introduced in \eqref{equiv_between_frames} reads:\footnote{To improve the clarity of the equations, the inherent dependence on $\ta_0$ in $S_{\mathrm{M}} = S_{\mathrm{M}_{\ta_0}}$ is explicitly displayed here.}
\begin{equation}\label{sss}
      S_{\mathrm{G},\tau-\ta_0} + S_{\mathrm{M},\tau_0} \simeq S_{\mathrm{EH}} + S_{\mathrm{M},\tau}\;,
\end{equation}
where $S_{\mathrm{M},\tau}$ is the $\TTb$-deformed matter action at the point $\ta$. On the other hand, $\tau_0$ is arbitrary, so we can promote it to a tunable parameter on the same standing as $\tau$. Relabelling the parameters in \eqref{sss}, we obtain
\begin{equation}
  S_{\mathrm{G},\ta_0} + S_{\mathrm{M},\tau-\ta_0}   \simeq S_{\mathrm{EH}} + S_{\mathrm{M},\tau}\;.
\end{equation}
This implies that 
\begin{equation}\label{general_flow_1}
      S_{\mathrm{G},\tau-\ta_0} + S_{\mathrm{M},\tau_0} \simeq S_{\mathrm{G},\ta_0} + S_{\mathrm{M},\tau-\ta_0}\;.
\end{equation}
For infinitesimal values of $\ta-\ta_0$, the dynamical equivalences \eqref{general_flow_1} induces the on-shell flow
\begin{equation}\label{flow_generic_1}
    \frac{\partial S_{\mathrm{G},\ta}}{\partial \ta} \simeq \frac{\partial S_{\mathrm{M},\ta}}{\partial \ta},
\end{equation}
where, by assumption, the expression on the right-hand side of equation \eqref{flow_generic_1} is governed by the $\mathrm{T}\overline{\mathrm{T}}$-like flow associated with the matter sector. Alternatively, flipping the sign of $\ta$ in the right-hand side of \eqref{flow_generic_1}, we may write
\begin{equation}\label{flow_generic_2}
    \frac{\partial }{\partial \ta}\left( S_{\mathrm{G},\ta} + S_{\mathrm{M},-\ta}\right)\simeq 0\;.
\end{equation}
On the other hand, regarding $\ta$ and $\ta_0$ as independent, and relabelling $\ta-\ta_0 \rightarrow \ta_1$ and $\ta_0 \rightarrow \ta_2$, we obtain
\begin{equation}\label{new_exchange}
 S_{\mathrm{G},\tau_1} + S_{\mathrm{M},\tau_2} \simeq S_{\mathrm{G},\ta_2} + S_{\mathrm{M},\ta_1}\;.   
\end{equation}
On-shell, the $\TTb$-like flow parameter and the characteristic scale of high-energy gravitational corrections are interchangeable. It is crucial to notice that, since our analysis is valid on the shell of the gravitational degrees of freedom, and recognising that the equations of motion of Ricci-based gravity theories exhibit a pronounced dependence on the specific source theory of matter to which they couple, the resulting stress tensor flow will strongly depend on the chosen matter sector. In the following sections, we will explicitly craft gravitational theories capable of accommodating the effects of trace-squared deformations, as well as $\TTb$-like deformations of Abelian gauge theories. The analytic results are presented for the $d=4$ case, even though, in principle, these techniques can be employed in arbitrary space-time dimensions. 
However, additional complications arise when extending the previous analysis beyond the four-dimensional setting. In section \ref{DimProblem}, we will explore why matters become more complex.
\subsection{Trace-squared deformations from Starobinsky gravity}
We begin our analysis by applying the results from section \ref{section3} to the Starobinsky model in four dimensions, which we introduced in \eqref{Starobinsky-4-d} as a simple prototype for $f(\cR)$ theories. Its Lagrangian density is given by
\begin{equation}
    \mathcal{L}_{\mathrm{Star},\kappa} = \frac{1}{2}\tr\left[\BR\right] + \frac{\kappa}{4}\tr\left[\BR\right]^2\,,
\end{equation}
where, as usual, $\tr\left[\BR\right]$ is seen as a function of some independent connection. Metric compatibility for such connection is achieved as in \eqref{h_explicit}, via the introduction of the auxiliary tensor
\begin{equation}\label{hstar_R}
   h_{\vargrav,\mu\nu} = 2\frac{\partial \cL_{\mathrm{Star},\vargrav}}{\partial \tr[\BR]}g_{\mu\nu} = \left(1+\vargrav \tr[\BR]\right)g_{\mu\nu}\,.
\end{equation}
We would now like to express $h_{\vargrav,\mu\nu}$ as a function of the stress-energy tensor of the matter sources in the model, and we do this by considering the equations of motion \eqref{traceofeoms}:
\begin{equation}\label{stareoms}
     2\frac{\partial \cL_{\mathrm{Star},\vargrav}}{\partial \tr[\BR]}\tr[\BR] - 4\cL_{\mathrm{Star},\vargrav} =-\tr[\BR]=\tr[\BT_{\ta_0}]\,.
\end{equation}
Plugging \eqref{stareoms} back into \eqref{hstar_R}, we get
\begin{equation}\label{h_star_to}
   h_{\vargrav,\mu\nu} = \left(1-\vargrav \tr[\BT_{\ta_0}]\right)g_{\mu\nu}\,.   
\end{equation}
On the other hand, from \eqref{corr-act}, we know that the Palatini frame action for some arbitrary matter sector can be written as
\begin{equation}
\begin{split}
  {S}_{\mathrm{M},\ta_0 + \vargrav}\left[h_{\vargrav,\mu\nu}, \Phi_I\right]&= \left.\left\{ S_\mathrm{M}\left[g_{\mu\nu},\Phi_I\right]+\int \mathrm{d}^4 \Bx \sqrt{-g}\,\left(\cL_{\mathrm{G},\vargrav}- \tr[\BR]\frac{\partial \cL_{\mathrm{G},\vargrav}}{\partial \tr[\BR]}\right)\right\}\right|_{g=g(h)}\\
  &= \left.\left\{ S_\mathrm{M}\left[g_{\mu\nu},\Phi_I\right]-\frac{\kappa}{4}\int \mathrm{d}^4 \Bx \sqrt{-g}\,\tr[\BR]^2\right\}\right|_{g=g(h)}\\
  &= \left.\left\{ S_\mathrm{M}\left[g_{\mu\nu},\Phi_I\right]-\frac{\kappa}{4}\int \mathrm{d}^4 \Bx \sqrt{-g}\,\tr[\BT_{\ta_0}]^2\right\}\right|_{g=g(h)}\,,
\end{split}
\end{equation}
where, going from the second to the third line, we used \eqref{stareoms}. Finally, setting $\kappa = a(\ta-\ta_0)$, we  find
\begin{equation}\label{star_dress}
 {S}_{\mathrm{M},\ta}\left[h_{\vargrav,\mu\nu}, \Phi_I\right]=\left.\left\{ S_\mathrm{M}\left[g_{\mu\nu},\Phi_I\right]-\frac{a(\ta-\ta_0)}{4}\int \mathrm{d}^4 \Bx \sqrt{-g}\,\tr[\BT_{\ta_0}]^2\right\}\right|_{g=g(h)}\,,   
\end{equation}
where the relation between $g_{\mu\nu}$ and $h_{\vargrav,\mu\nu}$ is specified by \eqref{h_star_to}, yielding, in terms of the new parameter $\ta$:
\begin{equation}\label{h_star_tau}
   h_{\ta,\mu\nu} = \left(1-a(\ta-\ta_0) \tr[\BT_{\ta_0}]\right)g_{\mu\nu}\,.   
\end{equation}
Equation \eqref{star_dress}, together with the metric deformation \eqref{h_star_tau}, define how the matter sector is modified in the Palatini frame of four-dimensional Starobinsky gravity. More importantly, they are the same equations describing the trace-squared dressing of arbitrary matter theories in $d=4$. This defines an exact duality between the dynamics of matter gravitating according to the Starobinsky model, and matter deformed by trace-squared deformations, as shown in equation \eqref{new_exchange}: 
\begin{equation}\label{ggggg}
  S_{\mathrm{Star},\ta_1} + S_{\mathrm{M},\ta_2} \simeq  S_{\mathrm{Star},\ta_2} + S_{\mathrm{M},\ta_1}\;.    
\end{equation}
Setting $\ta_1 = \ta$ and $\ta_2 = 0$, we see that \eqref{ggggg} reduces to
\begin{equation}\label{stareq}
    S_{\mathrm{Star},\tau} + S_{\mathrm{M}} \simeq S_{\mathrm{EH}} + S_{\mathrm{M},\tau}\;. 
\end{equation}
Matter theories in a Starobinsky-type universe behave as trace-squared deformed theories in the standard Einstein-Hilbert space-time. With the results from section \ref{section2} in mind, it is now intuitive to understand why, in a universe containing traceless matter -- such as, for example, pure Maxwell -- Starobinsky gravity has the same effects as General Relativity \cite{Delhom_2019}: the operator driving the dual stress tensor deformation automatically vanishes. Of course, the same statement holds for any seed matter theory with a traceless stress-energy tensor, such as the ModMax Lagrangian \eqref{modmax}. Notice also that the way trace-squared deformations affect the potential of scalar theories, such as in \eqref{def_sc_1}, reproduces the typical slow-roll suppression widely used in inflationary models \cite{Starobinsky:1979ty,Starobinsky:1980te,Vilenkin:1985md}. Infinitesimally, the gravity theory and the deformed matter theory share the same stress-tensor flow, according to \eqref{flow_generic_1}:
\begin{equation}
\frac{\partial S_{\mathrm{Star},\ta}}{\partial\ta} \simeq\frac{\partial S_{\mathrm{M},\ta}}{\partial\ta}=\frac{a}{4}\int \mathrm{d}^4\Bx \sqrt{-g}\,\tr\left[\BT_{\ta_0}\right]^2.    
\end{equation}
The Starobinsky/trace-squared duality serves as a very intuitive playground to understand how deformations of the gravity sector can be seen as deformations of the matter theory. With more complex RBG theories, the computations are more subtle, but the philosophy remains unchanged.

\subsubsection{The dimensionality problem}
\label{DimProblem}
As described in section \ref{section2}, trace-squared deformations in arbitrary dimensions correspond to metric deformations of the form
\begin{equation}\label{arb_tr2}
   h_{\ta,\mu\nu} = \left(1-a(\ta-\ta_0) \tr[\BT_{\ta_0}]\right)^{\frac{4}{d}}g_{\mu\nu}\,.   
\end{equation}
Since \eqref{arb_tr2} depends solely on the trace of $\BT_{\ta_0}$, it is natural to assume that the associated gravity theory should belong to the $f(\cR)$ family. Let us then consider some unspecified gravity Lagrangian $\cL_{\mathrm{G,\kappa}}$, and let us see what conditions must be imposed on it such that the metric deformation \eqref{arb_tr2} is realised on-shell. From \eqref{h_explicit}, we know that
\begin{equation}\label{h_explicit_2}
    h_{\vargrav,\mu\nu} = \left(2\frac{\partial \cL_{\mathrm{G},\vargrav}}{\partial \tr[\BR]}\right)^{\frac{2}{d-2}}g_{\mu\nu}\,.
\end{equation}
On the other hand, the $\tr[\BT_{\ta_0}]$ term in \eqref{arb_tr2} can be replaced with some functional of $\tr[\BR]$ using the equations of motion, so that (setting for convenience $a(\ta-\ta_0)=\vargrav$), one must have
\begin{equation}\label{diffL}
 \left(2\frac{\partial \cL_{\mathrm{G},\vargrav}}{\partial \tr[\BR]}\right)^{\frac{2}{d-2}} =  \left[1-\vargrav \left(  2\frac{\partial \cL_{\mathrm{G},\vargrav}}{\partial \tr[\BR]}\tr[\BR] - d\cL_{\mathrm{G},\vargrav}\right)\right]^{\frac{4}{d}}\,. 
\end{equation}
When $d=4$, the exponents on both sides automatically match: $2/(d-2)=4/d$. In arbitrary dimensions, this does not happen, and equation \eqref{diffL} cannot be analytically solved in general. This problem is not confined to $f(\cR)$ theories, and it is ultimately what renders it difficult to extend gravity/deformations dualities to generic space-time dimensions.

\subsection{Deformed Abelian gauge theories from EiBI gravity}
In this section, we consider four-dimensional Eddington-inspired Born-Infeld (EiBI) gravity, a Ricci-based gravity theory that will soon display profound links with the $\TTb$-like deformations of Abelian gauge theories discussed in section \ref{section2}. The $d=4$ EiBI action is given by
\begin{equation}\label{eibi}
S_{\mathrm{EiBI},\vargrav}\left[g_{\mu\nu},\Gamma^\lambda_{\mu\nu}\right] = \frac{1}{\vargrav} \int \mathrm{d}^4\Bx \left\{\sqrt{-\det\left[g_{\mu\nu} + \vargrav \mathcal{R}_{(\mu\nu)}(\Gamma) \right]}-\sqrt{-g}\right\}\,.    
\end{equation}
Computing the associated auxiliary metric, from \eqref{hforrbg} we obtain
\begin{equation}\label{hforeibi}
 h_{\vargrav,\mu\nu} = g_{\mu\nu}+\vargrav\cR_{(\mu\nu)}.   
\end{equation}
Notice that the determinant of  $h_{\vargrav,\mu\nu}$, defined in  \eqref{hforeibi}, appears explicitly in the EiBI action \eqref{eibi}, and this allows us to easily compute the equations of motion associated with the metric tensor as \cite{jimenez_2018}:
\begin{equation}\label{eoms:for:eibi}
    \sqrt{-h_\vargrav}\left(h_\vargrav^{-1}\right)^{\mu\nu} = \sqrt{-g}\left(g^{\mu\nu}-\vargrav T_{\ta_0}^{\mu\nu}\right)\,.
\end{equation}
We can now write the Palatini frame action for matter sources coupled to \eqref{eibi} in $d=4$:
\begin{equation}\label{corr-act_eibi}
\begin{split}
& \,\,\,\,S_{\mathrm{M},\ta_0 + \vargrav}\left[h_{\vargrav,\mu\nu},\Phi_I\right] \\=&\left.\left\{S_{\mathrm{M}}\left[g_{\mu\nu},\Phi_I\right]+\int \mathrm{d}^4\Bx\sqrt{-g}\,\left(\cL_{\mathrm{G},\vargrav}- g^{\mu\alpha}\cR_{(\alpha\nu)}\frac{\partial \cL_{\mathrm{G},\vargrav}}{\partial g^{\mu\beta}\cR_{(\beta\nu)}}\right)\right\}\right|_{g=g(h)}\!\! \\
=& \left.\left\{S_{\mathrm{M}}\left[g_{\mu\nu},\Phi_I\right]+\frac{1}{\vargrav}\int \mathrm{d}^4\Bx\,\left[\frac{1}{2}\sqrt{-h_\vargrav}\left(h_\vargrav^{-1}\right)^{\mu\nu}g_{\mu\nu}-\sqrt{-h_\vargrav}-\sqrt{-g}\right]\right\}\right|_{g=g(h)}\!\!\\
=&\left.\left\{S_{\mathrm{M}}\left[g_{\mu\nu},\Phi_I\right]+\frac{1}{\vargrav}\int \mathrm{d}^4\Bx\,\left[\frac{1}{2}\sqrt{-g}\left(4-\vargrav\tr[\BT_{\ta_0}]\right)-\sqrt{-h_\vargrav}-\sqrt{-g}\right]\right\}\right|_{g=g(h)}\!\!\\
=&\left.\left\{S_{\mathrm{M}}\left[g_{\mu\nu},\Phi_I\right]+\int \mathrm{d}^4\Bx\,\sqrt{-g} \left[\frac{1}{\vargrav}\left(1-\frac{\sqrt{-h_\vargrav}}{\sqrt{-g}}\right) -\frac{1}{2}\tr\left[\BT_{\ta_0}\right]\right]\right\}\right|_{g=g(h)}\!\!,
\end{split}
\end{equation}
where, going from the second to the third line, we made use of the equations of motion \eqref{eoms:for:eibi}. Using again \eqref{eoms:for:eibi}, and taking the determinant of both sides, we can eliminate the residual $h_\vargrav$ dependency in the last line, obtaining
\begin{equation}
\frac{\sqrt{-h_\vargrav}}{\sqrt{-g}} = \sqrt{\det\left[\mathbf{1}-\vargrav\BT_{\ta_0}\right]}\,.
\end{equation}
To pursue further simplifications, we need to give up full generality and introduce specific assumptions concerning the matter sector of the theory. For the current purposes, it is convenient to assume that the stress-energy tensor of the matter sector satisfies the degeneracy conditions discussed in \eqref{eq:Ddiag}. As pointed out before, such degeneracy is characteristic of Abelian gauge theories in $d=4$ (but not necessarily restricted to them). If $\BT_{\ta_0}$ admits only two independent eigenvalues, the following identity holds in the four-dimensional case:
\begin{equation}\label{useful_identity}
    1-\frac{1}{2}\vargrav\tr\left[ \BT_{\ta_0}\right]-\sqrt{\det\left[\mathbf{1}-\vargrav\BT_{\ta_0}\right]} = -\vargrav^2\sqrt{\det\left[ \BT_{\ta_0}\right]} \,.
\end{equation}
The validity of \eqref{useful_identity} is easily checked after diagonalising the stress-energy tensor. Making use of \eqref{useful_identity} in \eqref{corr-act_eibi}, we obtain
\begin{equation}
 S_{\mathrm{M},\ta_0 + \vargrav}\left[h_{\vargrav,\mu\nu},\Phi_I\right] =   \left.\left\{S_{\mathrm{M}}\left[g_{\mu\nu},\Phi_I\right]-\vargrav\int \mathrm{d}^4\Bx\,\sqrt{-g}\sqrt{\det\left[ \BT_{\ta_0}\right]} \right\}\right|_{g=g(h)} \!\!.
\end{equation}
Finally, under the degeneracy condition \eqref{eq:Ddiag}, it is possible to verify that the Einstein frame metric satisfies
\begin{equation}\label{inverse}
    h_{\vargrav,\mu\nu} = g_{\mu\nu} +\vargrav\left(T_{\ta_0,\mu\nu} - \frac{1}{2}\tr\left[\BT_{\ta_0}\right]g_{\mu\nu}\right).
\end{equation}
In fact,
\begin{equation}
\begin{split}
  &\left(h_{\vargrav}^{-1}\right)^{\mu\alpha}\left[g_{\alpha\nu} +\vargrav\left(T_{\ta_0,\alpha\nu} - \frac{1}{2}\tr\left[\BT_{\ta_0}\right]g_{\alpha\nu}\right)\right]\\
 =&\frac{\sqrt{-g}}{\sqrt{-h_\vargrav}}\left(g^{\mu\alpha}-\vargrav T^{\mu\alpha}_{\ta_0}\right)\left[g_{\alpha\nu} +\vargrav\left(T_{\ta_0,\alpha\nu} - \frac{1}{2}\tr\left[\BT_{\ta_0}\right]g_{\alpha\nu}\right)\right] \\
 =&\frac{\sqrt{-g}}{\sqrt{-h_\vargrav}} \left(\delta^\mu_\nu-\frac{1}{2}\vargrav\tr\left[\BT_{\ta_0}\right]\delta^\mu_\nu+\vargrav^2\sqrt{\det\left[\BT_{\ta_0}\right]}\delta^\mu_\nu\right)=\delta^\mu_\nu\,.
\end{split}
\end{equation}
As before, all is left to do is identify $\vargrav=\ta-\ta_0$: the matter action in the Palatini frame matches the $\TTb$-like dressing equation \eqref{need_this_later}, and the auxiliary metric \eqref{inverse} corresponds to the deformed metric \eqref{h4dgauge}. This shows that ordinary matter coupled to EiBI gravity shares the same dynamics of $\TTb$-like deformed matter which gravitates in accordance with the laws of General Relativity. In terms of flows, provided that the matter theory exhibits the desired degeneracy, we have
\begin{equation}\label{eibifl2}
    \frac{\partial S_{\mathrm{EiBI},\ta}}{\partial\ta} \simeq \frac{\partial S_{\mathrm{M},\ta}}{\partial\ta}=\int \mathrm{d}^4\Bx \sqrt{-g}\,\sqrt{\det\left[\BT_{\ta}\right]}\;.
\end{equation}
From the gravity point of view, it was observed in \cite{Delhom_2019} that coupling EiBI with Maxwell's theory yields identical dynamics to coupling General Relativity with Born-Infeld electromagnetism. However, when working from a purely gravitational perspective, calculations get messy when dealing with more complicated matter theories. In the $\TTb$-like framework, provided that the theory features the expected degeneracy, the following dynamical equivalence is always established:
\begin{equation}\label{ggggg2}
  S_{\mathrm{EiBI},\ta_1} + S_{\mathrm{M},\ta_2} \simeq  S_{\mathrm{EiBI},\ta_2} + S_{\mathrm{M},\ta_1}\;.    
\end{equation}
Choosing $\ta_1=\ta$ and $\ta_2 = 0$, we obtain
\begin{equation}\label{eibieq}
    S_{\mathrm{EiBI},\tau} + S_{\mathrm{M}} \simeq S_{\mathrm{EH}} + S_{\mathrm{M},\tau}\;. 
\end{equation}
For example, the computations in this section show that \textit{any} $U(1)$ gauge theory obeys \eqref{eibieq}. It is possible to express both Starobinsky and EiBI gravity as limiting cases of a single gravity theory: we shall explore this scenario in Appendix \ref{unif_frame}.
\section{Perturbative gravity flows}\label{section_final}
As we discussed in section \ref{section3}, the explicit four-dimensional correspondences presented in the previous sections originate from a more general setup in the context of RBGs. A natural question is whether a somehow weaker version of this setup can be extended to larger families of deformations, regardless of the existence of an exact dressing mechanism.  Even if we abandon the idea of the dressing mechanism, it may still be feasible to study more general deformations related to the stress-energy tensor perturbatively.  The idea is to focus on infinitesimal dualities, extending the family of gravitational on-shell flows introduced in \eqref{flow_generic_1} to 
\begin{equation}\label{general_flow_equation}
   \frac{\partial S_{\mathrm{G},\ta}}{\partial\ta} \simeq\frac{\partial S_{\mathrm{M},\ta}}{\partial\ta}=\int \mathrm{d}^d\Bx \sqrt{-g}\,\mathcal{O}_\ta\;,     
\end{equation}
where this time the choice of the deforming operator $\mathcal{O}_\ta$ is left completely arbitrary. It should be noted here that, despite the obvious similarity to previous results, the existence of stress tensor flows as \eqref{general_flow_equation} is a weaker property of field theories when compared to the class of full analytical dualities explored in previous sections. As a noteworthy example, we consider for the matter sector the four-dimensional Modified Nambu-Goto action:
\begin{equation}\label{mng}
     S_{\mathrm{MNG},\ta}\left[g_{\mu\nu},\phi\right] = \frac{1}{\ta}\int \mathrm{d}^4 \Bx \left\{\sqrt{-\det\left[g_{\mu\nu}+u\ta X_{\mu\nu}+v\ta X g_{\mu\nu}\right]}-\sqrt{-g}\right\} \, ,\quad u,\,v \in \mathbb{R}\;,
\end{equation}
where $X_{\mu\nu} = \partial_\mu\phi\partial_\nu\phi$, and $X$ simply denotes its trace. Expanding \eqref{mng} for small values of the coupling $\ta$, we obtain
\begin{equation}
    S_{\mathrm{MNG},\ta} \left[g_{\mu\nu},\phi\right]  = \frac{1}{2} \int \mathrm{d}^4 \Bx \sqrt{-g}\left(u+4v\right)X + O(\ta)\;,
\end{equation}
so that \eqref{mng} converges to the free bosonic action provided that $u+4v=1$. We henceforth impose this limit to consistently fix one of the free parameters of the theory. On the gravitational side, we consider the Modified Eddington-inspired Born-Infeld (MEiBI) action \cite{Chen_2016}
\begin{equation}\label{meibi}
S_{\mathrm{MEiBI},\vargrav}\left[g_{\mu\nu},\Gamma^\lambda_{\mu\nu}\right] = \frac{1}{\vargrav} \int \mathrm{d}^4\Bx \left\{\sqrt{-\det\left[g_{\mu\nu} + \alpha\vargrav \mathcal{R}_{(\mu\nu)}(\Gamma)+\beta\vargrav\tr[\BR]g_{\mu\nu} \right]}-\sqrt{-g}\right\}\,,    
\end{equation}
where $\alpha$ and $\beta$ are real parameters. Again, they are not fully independent, as one can see by expanding \eqref{meibi} in powers of $\vargrav$:
\begin{equation}    S_{\mathrm{MEiBI},\vargrav}\left[g_{\mu\nu},\Gamma^\lambda_{\mu\nu}\right] = \frac{1}{2} \int \mathrm{d}^4 \Bx \sqrt{-g}\left(\alpha+4\beta \right)\tr[\BR] + O(\vargrav)\;.
\end{equation}
As we did for \eqref{mng}, we fix $\alpha+4\beta=1$ to recover the Einstein-Hilbert action at the leading order. It is now easy to show that, assuming
\begin{equation}\label{sys3}
    \begin{cases}
        u = \alpha\,,\\
        v = \beta\,,\\
        \alpha\, \cR_{(\mu\nu)}+\beta \,\tr\left[\BR\right] g_{\mu\nu} = -u\,X_{\mu\nu} -vXg_{\mu\nu}\;.
    \end{cases}
\end{equation}
the following on-shell flow is realised:
\begin{equation}
    \frac{\partial}{\partial \ta} \left( S_{\mathrm{MEIBI},\ta} + S_{\mathrm{MNG},-\ta}\right) \simeq 0\;.
\end{equation}
Taking the trace of the last equation in \eqref{sys3}, and using $u + 4 v = 1$, we obtain
\begin{equation}\label{einst_eoms}
    \cR_{(\mu\nu)} = - X_{\mu\nu}.
\end{equation}
On the other hand, switching to matrix notation, explicit computation shows that
\begin{equation}\label{X_value}
    \mathbf{X} =  \frac{1}{2}\tr\left[\BT_{0}\right]\mathbf{1}-\BT_{0}\;,
\end{equation}
where $T_{0,\mu\nu}$ is the stress-energy tensor associated to the modified Nambu-Goto action in $\ta=0$, i.e., to the free bosonic action
\begin{equation}\label{fb3}
    S_{\mathrm{FB}} \left[g_{\mu\nu},\phi\right] =\frac{1}{2} \int \mathrm{d}^4 \Bx \sqrt{-g}X\,. 
\end{equation}
We immediately recognise that \eqref{einst_eoms} and \eqref{X_value} are nothing more than Einstein's field equations from four-dimensional General Relativity, provided that the matter sector is described by \eqref{fb3}. This tells us that the following dynamical equivalence must be satisfied:
\begin{equation}\label{mng_reframing}
S_{\mathrm{MEiBI},\ta} + S_{\mathrm{MNG},-\ta} \simeq S_{\mathrm{EH}} + S_{\mathrm{FB}}\;.   
\end{equation}
Notice that, setting $v=0$ in \eqref{mng}, the matter action reduces to the Nambu-Goto action in the static gauge:
\begin{equation}\label{ng}
    S_{\mathrm{NG},\ta}\left[g_{\mu\nu},\phi\right] = \frac{1}{\ta}\int \mathrm{d}^4 \Bx \left\{\sqrt{-\det\left[g_{\mu\nu}+\ta X_{\mu\nu}\right]}-\sqrt{-g}\right\}. 
\end{equation}
In turn, the dynamical equivalence is realised by fixing $\beta=0$ in the gravitational sector, and the theory reduces to EiBI gravity. Accordingly, equation \eqref{mng_reframing} yields
\begin{equation}\label{ng_reframing}
S_{\mathrm{EiBI},\ta} + S_{\mathrm{NG},-\ta} \simeq S_{\mathrm{EH}} + S_{\mathrm{FB}}\;.  
\end{equation}
This last outcome aligns with the findings reported in \cite{Afonso:2018hyj}. As shown in \cite{Ferko:2023sps}, the emergence of Nambu-Goto-like theories as a response to stress tensor deformations is generally associated with families of operators which are way more intricate than those explicitly examined in this paper. However, it is interesting that we can generate deformations of classical field theories by coupling their corresponding Lagrangians to modified theories of gravity, using the concept of Lagrangian flow equation. In this specific case, since the right-hand side of \eqref{ng_reframing} does not depend on $\tau$, the analogue of \eqref{general_flow_equation} is the on-shell flow
\begin{equation}
   \frac{\partial S_{\mathrm{G},\ta}}{\partial\ta} \simeq\frac{\partial S_{\mathrm{NG},\ta}}{\partial\ta}=\int \mathrm{d}^4\Bx \sqrt{-g}\,\mathcal{O}_\ta^\mathfrak{R}\,,       
\end{equation}
where the deforming operator $\mathcal{O}_\ta^\mathfrak{R}$ is defined by \cite{Ferko:2023sps}
\begin{equation}
\mathcal{O}_\ta^\mathfrak{R} = \frac{1}{16}\tr\left[\BT_\ta\right]^2 -\frac{1}{8}\tr\left[\BT_\ta^2\right]-\frac{1}{8}\tr\left[\BT_\ta\right]\sqrt{\frac{1}{3}\left(\tr\left[\BT_\ta^2\right]-\frac{1}{4}\tr\left[\BT_\ta\right]^2\right)}\;.
\end{equation}
\section{Conclusions}
In this paper, we examined explicit examples illustrating the connection between $\TTb$-like deformations in dimensions greater than two and Palatini's theories of gravity. This correspondence is associated with the dynamical equivalence \eqref{eq:conj} and explored through the $\TTb$-like dressing mechanism discussed in section 
\ref{sec:dressing} and the on-shell flow equation \eqref{flow_generic_1}. We introduced a broadly applicable scheme for investigating this connection and confirmed that the equivalence \eqref{eq:conj} arises from a process of reframing within Ricci-based gravity theories coupled to matter. While we discussed several physically interesting models in $d = 4$, there are aspects touched on in this article that merit further investigation. For example, a potential avenue for future research might correspond to the generalisation of the analysis of $U(1)$ gauge theories presented here to encompass non-Abelian Yang-Mills models and exploring the Root-$\TTb$ class of marginal deformations \cite{Conti_2022, Ferko:2022cix, Ferko:2023ruw, Babaei-Aghbolagh:2022uij, Babaei-Aghbolagh:2022leo}. In this respect, it would be important to understand whether it is possible to compute the $\TTb$ deformed metric and Lagrangian associated with non-Abelian gauge theories (possibly coupled to gravity) by relying on the method of characteristics \cite{Hou:2022csf}. Furthermore, it was demonstrated in \cite{Floss:2023nod} that integrating a massive graviton leads to the most general duality-invariant vector interactions in four-dimensional spacetime, with the specific case of Born-Infeld theory arising from ghost-free de Rham-Gabadadze-Tolley (dRGT) massive gravity \cite{deRham:2010kj}. Therefore, another potentially fruitful study might be conducted based on the similarity between the natural bimetric structure of dRGT gravity and the Einstein-frame representation of Eddington-inspired Born-Infeld gravity.
\medskip

\noindent{\bf Acknowledgments --}
We thank Riccardo Conti, Jacopo Romano and  Nicolò Brizio for their insightful discussions and valuable contributions to our previous collaborations on related research projects. We thank Christian Ferko, Gabriele Tartaglino-Mazzucchelli, Dmitri Sorokin, Arkady Tseytlin, Anton Pribytok and Jue Hou for their valuable insights and comments. We also thank Christian Ferko for sharing notes which independently observed the connection between stress tensor deformations and Eddington-inspired Born-Infeld gravity, to leading order in the deformation parameter. We are grateful to Sylvain Fichet for his enriching remarks, which led to the integration of Appendix \ref{sec:dofs} in the updated version of this paper.

\medskip

This project received partial support from the INFN project SFT and the PRIN Project No. 2022ABPBEY, with the title ``Understanding quantum field theory through its deformations''. Roberto Tateo expresses gratitude to the participants of the MATRIX Research Program ``New Deformations of Quantum Field and Gravity'' for the stimulating atmosphere and discussions, and to the Mathematical Research Institute MATRIX and the Sydney Mathematical Research Institute for the invitation and financial support during his Australian visit during the last stages of this project.

\appendix

\section{Counting degrees of freedom}\label{sec:dofs}

For scalar-tensor theories of gravity, in the so-called Jordan frame, matter fields are minimally coupled to the metric. Conversely, in the Einstein frame, gravity is described by the Einstein-Hilbert term, and matter fields couple to a conformally deformed metric, with the conformal factor being dependent on some scalar field. In the context of RBGs, a similar scenario unfolds, albeit with a crucial distinction: there are no additional propagating degrees of freedom \cite{jimenez_2018}. As an illustrative example, we turn our attention to the class of $f\left(\cR\right)$ theories in four-dimensional space-time. We start by introducing an auxiliary scalar field $\phi$ -- usually called the \textit{dilaton field} -- defined by
\begin{equation}
   \phi = 2  \frac{\partial \cL_{\mathrm{G},\vargrav}}{\partial \tr\left[\BR\right]}\,.
\end{equation}
In terms of the dilaton, the conformal rescaling of the metric introduced in equation \eqref{sigmaos} can be written as
\begin{equation}
    h_{\vargrav,\mu\nu} = \phi\, g_{\mu\nu}\;.
\end{equation}
Moreover, notice that equation \eqref{rhs2} becomes
\begin{equation}\label{scalar-tensor1}
\widehat{S}_\vargrav = \frac12\int \mathrm{d}^4 \Bx \sqrt{-g}\left(\phi \, \tr\left[\BR\right] - V(\phi)\right)+S_\mathrm{M}\left[g_{\mu\nu},\Phi_I\right]\,,
\end{equation}
where we introduced the dilaton potential
\begin{equation}
    V(\phi) := \phi\,\tr\left[\BR\right] -  2\cL_{\mathrm{G},\vargrav}\,.
\end{equation}
As a simple example, consider the $d=4$ Starobinsky Lagrangian
\begin{equation}
    \mathcal{L}_{\mathrm{G},\kappa} = \frac{1}{2}\tr\left[\BR\right] + \frac{\kappa}{4}\tr\left[\BR\right]^2.
\end{equation}
The associated dilaton potential is then
\begin{equation}
    V(\phi) = \frac{4}{\kappa} (\phi -1)^2\,.
\end{equation}
If we define the Christoffel symbols of the metric $g_{\mu\nu}$ as
\begin{equation}
\Theta^\lambda_{\mu\nu} =\frac12 g^{\lambda\alpha} \left(\partial_\nu g_{\mu\alpha} + \partial_\mu g_{\alpha\nu} - \partial_\alpha g_{\mu\nu}\right), 
\end{equation}
and introduce the associated scalar curvature
\begin{equation}
    R(g) = g^{\mu\nu}\left( \partial_\alpha \Theta_{\nu \mu}^\alpha-\partial_\nu \Theta_{\alpha \mu}^\alpha+\Theta_{\alpha \beta}^\alpha \Theta_{\nu \mu}^\beta-\Theta_{\nu \beta}^\alpha \Theta_{\alpha \mu}^\beta\right),
\end{equation}
it is possible to rewrite equation \eqref{scalar-tensor1} as \cite{Faraoni:2008bu}
\begin{equation}
\widehat{S}_\vargrav = \frac12\int \mathrm{d}^4 \Bx \sqrt{-g}\left[\phi R(g)-\frac{\omega}{\phi}\left(\partial_\mu \phi \partial^\mu \phi\right)-V(\phi)\right]+S_\mathrm{M}\left[g_{\mu\nu},\Phi_I\right]\,,\quad \omega = -\frac{3}{2}\,.
\end{equation}
Hence, the reframed action for $f\left(\cR\right)$ gravity turns out to be equivalent to a Brans-Dicke gravity theory \cite{Brans:1961sx} with coupling parameter $\omega = -\frac{3}{2}\,$. In the
Brans-Dicke theory, the scalar field $\phi$ is governed by the equation
\begin{equation}\label{boxphi}
    (3+2 \omega) \square \phi+2 V(\phi)-\phi \frac{d V}{d \phi}= \tr\left[\BT_{\ta_0}\right],
\end{equation}
which, using $\omega = -\frac{3}{2}\,$, reduces to
\begin{equation}\label{nobox}
2 V(\phi)-\phi \frac{d V}{d \phi}=\tr\left[\BT_{\ta_0}\right].
\end{equation}
Notice that the dynamical field equation \eqref{boxphi} for the Brans-Dicke scalar degenerates into the algebraic identity \eqref{nobox}. In particular, the lack of dynamics for the dilaton field explicitly demonstrates that the change of frame does not introduce any new propagating degrees of freedom.
It is intriguing to observe the notable resemblance between the features of two-dimensional JT gravity \cite{Dubovsky_2017} and the emergence of the dilaton field in the Brans-Dicke representations of $f\left(\cR\right)$ theories.

\section{A possible unified framework for EiBI and Starobinsky gravity}\label{unif_frame}
In \cite{Chen_2016}, the $d=4$ Modified Eddington-inspired Born-Infeld (MEiBI) gravity action was introduced as  the most general action constructed from a rank-two tensor that contains up to first-order terms in curvature:
\begin{equation}\label{meibiappendix}    S_{\mathrm{MEiBI},\vargrav}\left[g_{\mu\nu},\Gamma^\lambda_{\mu\nu}\right] = \frac{1}{\vargrav} \int \mathrm{d}^4\Bx \left\{\sqrt{-\det\left[g_{\mu\nu} + \vargrav \mathcal{P}_{\mu\nu}(\Gamma) \right]}-\sqrt{-g}\right\}\,.
\end{equation}
Here $\mathcal{P}_{\mu\nu}$ is a rank-$2$ covariant tensor constructed from the symmetric part of the Ricci tensor $\cR_{(\mu\nu)}$ in the Palatini formalism:
\begin{equation}\label{Ptensor}
    \mathcal{P}_{\mu\nu} := \alpha \cR_{(\mu\nu)} + \beta \tr\left[ \mathbf{R}\right]g_{\mu\nu},\quad \alpha,\,\beta \in \mathbb{R}\;.
\end{equation}
The MEiBI action was briefly introduced in section \ref{section_final} when discussing generalisations of gravity flows to Nambu-Goto-like matter Lagrangians. Notice that the action contains a mass scale induced by the dimensional parameter of the theory, given by
\begin{equation}
 m_\mathrm{G}^{-2} = \vargrav\,.  
\end{equation}
If we insist that in the infrared limit -- i.e., for large enough values of $m_\mathrm{G}$ -- the action \eqref{meibiappendix} should reduce to General Relativity, the free parameters $\alpha$ and $\beta$ are found not to be entirely independent. By expanding the MEiBI action for small values of $\vargrav$, we can see that
\begin{equation}
    S_{\mathrm{MEiBI},\vargrav} = \frac{1}{2} \int \mathrm{d}^4\Bx \sqrt{-g} \left(\alpha + 4\beta\right)\tr\left[ \mathbf{R}\right] + O(\vargrav)\;.
\end{equation}
Therefore, if we set 
\begin{equation}\label{ab}
    \alpha + 4\beta = 1\,,
\end{equation}
at lower orders, we retrieve the conventional Einstein-Hilbert action. Then, the action \eqref{meibiappendix} represents a one-parameter family of gravity theories, labelled by $\beta\in\mathbb{R}$. It is straightforward to notice that fixing $\beta = 0$ -- and consequently $\alpha = 1$ -- the action of MEiBI gravity reduces to the EiBI action \eqref{eibi}. Less obviously, by setting $\alpha=0$ and $\beta =1/4$, we recover the Starobinsky action \eqref{Starobinsky-4-d}. To see this, notice that the integrand in \eqref{meibiappendix} becomes
\begin{equation}\label{starfrommeibi}
\begin{split}
 \frac{1}{\vargrav}\left\{\sqrt{-\det\left[g_{\mu\nu} + \frac{\vargrav}{4}\tr\left[\BR\right]g_{\mu\nu} \right]}-\sqrt{-g}  \right\}=   \frac{1}{\vargrav}\sqrt{-g} \left\{\left(1+\frac{\vargrav}{4}\tr\left[\BR\right]\right)^{2}-1\right\}.    
\end{split}
\end{equation}
The coefficient of the $\tr\left[\BR\right]^{2}$ term in \eqref{starfrommeibi} differs by a factor of $1/4$ if compared with the action introduced in \eqref{Starobinsky-4-d}, by the full equivalence can be restored through a rescaling of $\vargrav$. It is fascinating to observe that EiBI gravity and Starobinsky gravity, both intimately related to $\TTb$-like deformations in four-dimensional space-time, can be interpreted as originating from the more general framework of MEiBI theories. In this appendix, we study the equations of motion associated with \eqref{meibiappendix}, and we apply the reframing procedure discussed in section \ref{section3}. For the sake of full generality, we extend the MEiBI action to arbitrary $d$ dimensional space-time, and account for the presence of a non-vanishing cosmological constant as follows:
\begin{equation}\label{meibid}    S_{\mathrm{MEiBI},\vargrav}\left[g_{\mu\nu},\Gamma^\lambda_{\mu\nu}\right] = \frac{1}{\vargrav} \int \mathrm{d}^d\Bx \left\{\sqrt{-\det\left[g_{\mu\nu} + \vargrav \mathcal{P}_{\mu\nu}(\Gamma) \right]}-\lambda\sqrt{-g}\right\}\,.
\end{equation}
Expanding, perturbatively,  \eqref{meibid} for small values of $\vargrav$, we obtain
\begin{equation}
    S_{\mathrm{MEiBI},\vargrav} = \frac{1}{2} \int \mathrm{d}^d\Bx \sqrt{-g} \left[\left(\alpha + d\beta\right)\tr\left[ \mathbf{R}\right]-2\left(\frac{\lambda-1}{\vargrav}\right)\right] + O(\vargrav)\;.
\end{equation}
Therefore, we identify $\alpha+d\beta=1$, and we introduce an effective cosmological constant
 \begin{equation}\label{Lambda}
     \Lambda = \frac{\lambda-1}{\vargrav}\,.
 \end{equation}
 The next step involves incorporating matter into the action, which we do via the minimal coupling prescription:
 \begin{equation}\label{meibi+matter}
      S_{\vargrav} \left[g_{\mu\nu}, \Gamma^\lambda_{\mu\nu} ,\Phi_I\right]=  S_{\mathrm{MEiBI},\vargrav}\left[g_{\mu\nu}, \Gamma^\lambda_{\mu\nu}\right] +  S_\mathrm{M}\left[g_{\mu\nu}, \Phi_I\right]\,.
 \end{equation}
Once again, we assume that $\Gamma^\lambda_{\mu\nu}$ does not enter the matter action $S_\mathrm{M}$. To obtain the equations of motion for \eqref{meibi+matter}, it is quite convenient to introduce the tensor
\begin{equation}\label{ppp}
    p_{\mu\nu} := g_{\mu\nu} + \vargrav \cP_{\mu\nu}\;.
\end{equation}
The variation of \eqref{meibi+matter} with respect to the metric $g_{\mu\nu}$ yields the field equations \cite{Chen_2016}
\begin{equation}\label{field_equations_meibi}
\left(p^{-1}\right)^{\mu\nu}\left(1+\beta\vargrav\cR_{(\alpha\beta)}g^{\alpha\beta}\right)-\beta\vargrav \left(p^{-1}\right)^{\alpha\beta}g_{\alpha\beta}g^{\mu\rho}g^{\nu\sigma}\cR_{(\rho\sigma)} = \gamma \left(\lambda g^{\mu\nu}-\vargrav T_{\ta_0}^{\mu\nu}\right),
\end{equation}
where $\left(p^{-1}\right)^{\mu\nu}$ denotes the inverse of $p_{\mu\nu}$, and we introduced the ratio
\begin{equation}
    \gamma := \frac{\sqrt{-g}}{\sqrt{-p}}\;,\qquad p := \det\left[p_{\mu\nu}\right].
\end{equation}
Because the matter sector is assumed to be covariantly coupled to the metric $g_{\mu\nu}$ only, the energy-momentum tensor is conserved
as in General Relativity. On the other hand, the variation of \eqref{meibi+matter} with respect to the connection $\Gamma^{\lambda}_{\mu\nu}$ yields,
\begin{equation}\label{gamma_variation}
    \nabla_\mu\left\{\sqrt{-p}\left[\alpha \left(p^{-1}\right)^{\mu\nu}+ \beta \left(p^{-1}\right)^{\alpha\beta}g_{\alpha\beta}g^{\mu\nu}\right]\right\}=0,
\end{equation}
where the covariant derivative is defined with respect to the connection $\Gamma^{\lambda}_{\mu\nu}$. Equation \eqref{gamma_variation} implies the existence of an auxiliary metric tensor $h_{\vargrav,\mu\nu}$, defined by its inverse through
\begin{equation}\label{hp}
\sqrt{-h_\vargrav}\left(h_\vargrav^{-1}\right)^{\mu\nu}:=\sqrt{-p}\left[\alpha \left(p^{-1}\right)^{\mu\nu}+ \beta \left(p^{-1}\right)^{\alpha\beta}g_{\alpha\beta}g^{\mu\nu}\right],  \quad h := \det\left[h_{\mu\nu}\right],
\end{equation}
such that $\Gamma^\lambda_{\mu\nu}$ is $h_\vargrav$-compatible, meaning that
\begin{equation}\label{h_compatibility}
  \Gamma^\lambda_{\mu\nu} =\frac12 \left(h_\vargrav^{-1}\right)^{\lambda\alpha} \left(\partial_\nu h_{\vargrav,\mu\alpha} + \partial_\mu h_{\vargrav,\alpha\nu} - \partial_\alpha h_{\vargrav,\mu\nu}\right).
\end{equation}
It is worth noting that, from \eqref{hp}, we can calculate 
\begin{equation}
 \sqrt{-h_\vargrav} = \sqrt{-p}\, \det\left[\alpha \delta^{\mu}_{\nu}+ \beta \left(p^{-1}\right)^{\alpha\beta}g_{\alpha\beta}g^{\mu\sigma}p_{\sigma\nu}\right]^{\frac{1}{d-2}}:=f\left(\alpha,\beta\right)\sqrt{-p}\;.
\end{equation}
The dynamical equivalence discussed in section \ref{section3} can be achieved through the introduction of the auxiliary functional
\begin{equation}\label{meibiequiv}
 \begin{split}
\widehat{S}_{\mathrm{MEiBI},\vargrav}\left[g_{\mu\nu},h_{\vargrav,\mu\nu},\Gamma^\lambda_{\mu\nu}\right] &= \frac{1}{2}\int \mathrm{d}^d \Bx \sqrt{-h_\vargrav}\left(h_\vargrav^{-1}\right)^{\mu\nu}\cR_{(\mu\nu)}\\&+\frac{1}{\vargrav}\int \mathrm{d}^d\Bx \left(\frac{1}{2}\sqrt{-h_\vargrav}\left(h_\vargrav^{-1}\right)^{\mu\nu}g_{\mu\nu}-\frac{d-2}{2f(\alpha,\beta)}\sqrt{-h}-\lambda\sqrt{-g}\right),    
 \end{split}\end{equation}
where $h_{\vargrav,\mu\nu}$ is now some independent field. The notational overlap with the previously defined auxiliary metric is, once again, not accidental since, imposing the equation of motion for $h$ and re-expressing the on-shell value $h_{\vargrav,\mu\nu}^\star$ in terms of $p_{\mu\nu}$ via the definition \eqref{hp}, we find
\begin{equation}
\left.p_{\mu\nu}\right|_{h_\vargrav^\star} =  g_{\mu\nu} + \cP_{\mu\nu}\;.
\end{equation}
Substituting this result back into \eqref{meibiequiv}, we obtain
\begin{equation} \widehat{S}_{\mathrm{MEiBI},\vargrav}\left[g_{\mu\nu},h^\star_{\vargrav,\mu\nu},\Gamma^\lambda_{\mu\nu}\right] =  \frac{1}{\vargrav} \int \mathrm{d}^d\Bx \left\{\sqrt{-\det\left[g_{\mu\nu} + \vargrav \mathcal{P}_{\mu\nu}(\Gamma) \right]}-\lambda\sqrt{-g}\right\}, 
\end{equation}
which proves the dynamical equivalence between \eqref{meibi} and \eqref{meibiequiv}. As expected, the first term appearing on the right-hand side of \eqref{meibiequiv} is nothing more than the Einstein-Hilbert action, written in terms of the field $h_{\vargrav,\mu\nu}$. For later convenience, we define the remaining piece of \eqref{meibiequiv} as
\begin{equation}
    S_{+,\vargrav} \left[g_{\mu\nu},h_{\vargrav,\mu\nu}\right]:= \frac{1}{\vargrav}\int \mathrm{d}^d\Bx \left(\frac{1}{2}\sqrt{-h_\vargrav}\left(h_\vargrav^{-1}\right)^{\mu\nu}g_{\mu\nu}-\frac{d-2}{2f(\alpha,\beta)}\sqrt{-h_\vargrav}-\lambda\sqrt{-g}\right),
\end{equation}
so that
\begin{equation}
\widehat{S}_{\mathrm{MEiBI},\vargrav}\left[g_{\mu\nu},h_{\vargrav,\mu\nu},\Gamma^\lambda_{\mu\nu}\right] =   {S}_{\mathrm{EH}} \left[h_{\vargrav,\mu\nu},\Gamma^\lambda_{\mu\nu}\right] +  S_{+,\vargrav} \left[g_{\mu\nu},h_{\vargrav,\mu\nu}\right]\,.
\end{equation}
Coupling $\widehat{S}_{\mathrm{MEiBI},\vargrav}$ with matter produces the full action
\begin{equation}
\begin{split}
     \widehat{S}_{\vargrav} \left[g_{\mu\nu}, h_{\vargrav,\mu\nu},\Gamma^\lambda_{\mu\nu} ,\Phi_I\right]&=  \widehat{S}_{\mathrm{MEiBI},\vargrav}\left[g_{\mu\nu}, h_{\vargrav,\mu\nu},\Gamma^\lambda_{\mu\nu}\right] +  S_\mathrm{M}\left[g_{\mu\nu}, \Phi_I\right]\\
     &={S}_{\mathrm{EH}} \left[h_{\vargrav,\mu\nu},\Gamma^\lambda_{\mu\nu}\right] +  S_{+,\vargrav} \left[g_{\mu\nu},h_{\vargrav,\mu\nu}\right]+  S_\mathrm{M}\left[g_{\mu\nu}, \Phi_I\right]\,.    
\end{split}
\end{equation}
We can integrate out $g_{\mu\nu}$, which is now a purely algebraic field, and substitute its on-shell value $g^\star(h_\vargrav)$ as a function of the auxiliary metric $h_{\vargrav,\mu\nu}$. This enables the definition of a modified matter sector in terms of $h_{\vargrav,\mu\nu}$ alone as
\begin{equation}\label{modified_matter}
S_{\mathrm{M},\ta_0 + \vargrav}\left[h_{\vargrav,\mu\nu}, \Phi_I\right] := \left.\left\{S_\mathrm{M}\left[g_{\mu\nu}, \Phi_I\right]+S_{+,\vargrav} \left[g_{\mu\nu},h_{\vargrav,\mu\nu}\right] \right\}\right|_{g=g(h)},
\end{equation}
which is non-trivially coupled to the usual action from General Relativity. Notice that, since using the definition \eqref{hp}, together with the constraint \eqref{ab}, we have
\begin{equation}
\begin{split}
    \sqrt{-h_\vargrav}\left(h_\vargrav^{-1}\right)^{\mu\nu}g_{\mu\nu} &= \sqrt{-p}\left[\alpha \left(p^{-1}\right)^{\mu\nu}+ \beta \left(p^{-1}\right)^{\alpha\beta}g_{\alpha\beta}g^{\mu\nu}\right]g_{\mu\nu}   \\
    &= \sqrt{-p}\left(\alpha +d\beta\right)\left(p^{-1}\right)^{\mu\nu}g_{\mu\nu}  \\
    &= \sqrt{-p}\left(p^{-1}\right)^{\mu\nu}g_{\mu\nu}\;,    
\end{split}
\end{equation}
and since
\begin{equation}
 \frac{d-2}{2f(\alpha,\beta)}\sqrt{-h_\vargrav} = \frac{d-2}{2}\sqrt{-p}\;,
\end{equation}
we can write
\begin{equation}\label{meibiequiv2}
 \begin{split}
 S_{+,\vargrav} &=\frac{1}{\vargrav}\int \mathrm{d}^d\Bx \left(\frac{1}{2}\sqrt{-p}\left(p^{-1}\right)^{\mu\nu}g_{\mu\nu}-\frac{d-2}{2}\sqrt{-p}-\lambda\sqrt{-g}\right) \\
&=\frac{1}{\vargrav}\int \mathrm{d}^d\Bx\frac{\sqrt{-g}}{\gamma} \left(\frac{1}{2}\left(p^{-1}\right)^{\mu\nu}g_{\mu\nu}-\frac{d-2}{2}-\lambda\gamma\right).\\
 \end{split}\end{equation}
 Taking the trace of the field equations \eqref{field_equations_meibi} we obtain
 \begin{equation}\label{trace_of_eoms}
 \begin{split} \left(p^{-1}\right)^{\mu\nu}g_{\mu\nu} & = \left(p^{-1}\right)^{\mu\nu}\left(1+\beta\vargrav\cR_{(\alpha\beta)}g^{\alpha\beta}\right)g_{\mu\nu}-\beta\vargrav \left(p^{-1}\right)^{\alpha\beta}g_{\alpha\beta}g^{\mu\rho}g^{\nu\sigma}\cR_{(\rho\sigma)}g_{\mu\nu}  \\
  &=\gamma \left(\lambda g^{\mu\nu}-\vargrav T^{\mu\nu}\right)g_{\mu\nu}
  =\gamma \left(\lambda d-\vargrav \tr[\BT_{\ta_0}]\right),
 \end{split}
 \end{equation}
 and using these results in \eqref{meibiequiv2} we finally get
 \begin{equation}\label{meibiequiv3}
 \begin{split}
S_{+,\vargrav} =\int \mathrm{d}^d\Bx\sqrt{-g} \left( \frac{(d-2) (\lambda \gamma-1)}{2\gamma\vargrav}-\frac{1}{2}\tr\left[\BT_{\ta_0}\right]\right),
 \end{split}
 \end{equation}
thus the modified matter sector will be determined by
\begin{equation}\label{bbb}
   \left.\left\{S_\mathrm{M}\left[g_{\mu\nu}, \Phi_I\right] + \int \mathrm{d}^d\Bx\sqrt{-g} \left( \frac{(d-2) (\lambda \gamma-1)}{2\gamma\vargrav}-\frac{1}{2}\tr\left[\BT_{\ta_0}\right]\right)\right\}\right|_{g=g(h)}= S_{\mathrm{M},\ta_0 + \vargrav}\left[h_{\vargrav,\mu\nu}, \Phi_I\right] \,.
\end{equation}

\newpage 

\bibliography{biblio3}

\end{document}